\documentclass[journal,12pt,onecolumn,draftclsnofoot]{IEEEtran}
\ifCLASSINFOpdf

\else

\fi

\usepackage{color}
\usepackage{comment}
\usepackage{graphicx,tabularx,array,amsmath,amsthm,thmtools,subcaption}
\usepackage{enumitem}
\usepackage{mathtools}

\usepackage{caption}
\usepackage{amsfonts}
\usepackage{bm}
\usepackage{bbm}
\usepackage{makecell}
\usepackage{multirow}
 \usepackage{amssymb}
\usepackage{amsthm}
\usepackage{txfonts}
\usepackage[T1]{fontenc}
\usepackage[scr=dutchcal]{mathalfa}
\let\mathscr\mathbscr

\usepackage{cite}
\usepackage{amsmath}
\usepackage{booktabs}
\usepackage{cuted}
\usepackage{algorithm}
\usepackage{algorithmic}

\def\gbf{\mathbf{g}}
\newcommand{\bff}[1]{\mathbf{#1}}

\allowdisplaybreaks 

\declaretheoremstyle[headfont=\bfseries, 
    bodyfont=\normalfont]{normalhead}

\newtheorem{Theorem}{Theorem}

\newtheorem{Lemma}{Lemma}

\DeclareMathOperator*{\argmax}{argmax}

\newcommand{\E}[1]{\mathrm{E}\left[#1\right]}
\newcommand{\Var}[1]{\mathrm{Var}\left[#1\right]}

\begin{document}



\title{Multi-point Coordination in Massive MIMO Systems with Sectorized Antennas}




 \author{
\IEEEauthorblockN{Shahram Shahsavari, Mehrdad Nosrati, Parisa Hassanzadeh,
Alexei Ashikhmin, Thomas L. Marzetta, and Elza Erkip\\}
\thanks{S. Shahsavari and P. Hassanzadeh are with Huawei Technologies and J.P. Morgan, respectively. They have been with ECE Dept. of New York University, Brooklyn, NY when working on this paper. \{shahram.shahsavari, ph990\}@nyu.edu.} 
\thanks{M. Nosrati is with SpaceX, WA. He has been with ECE Dept. of Stevens Institute of Technology when working on this paper. mnosrati@stevens.edu.}
\thanks{T.~L.~Marzetta, and E. Erkip are with the ECE Dept. of New York University, Brooklyn, NY. \{tom.marzetta, elza\}@nyu.edu.}
\thanks{A. Ashikhmin is with Bell Labs, Nokia, Murray Hill, NJ. alexei.ashikhmin@nokia-bell-labs.com.}

}

\maketitle

\begin{abstract}
Non-cooperative cellular massive MIMO, combined with power control, is known to lead to significant improvements in per-user throughput compared with conventional LTE technology. In this paper, we investigate further refinements to massive MIMO, first, in the form of three-fold sectorization, and second, coordinated multi-point operation (with and without sectorization), in which the three base stations cooperate in the joint service of their users. For these scenarios, we analyze the downlink performance for both maximum-ratio and zero-forcing precoding and derive closed-form lower-bound expressions on the achievable rate of the users. These expressions are then used to formulate power optimization problems with two throughput fairness criteria: \textit{i}) network-wide max-min fairness, and \textit{ii}) per-cell max-min fairness. Furthermore, we provide centralized and decentralized power control strategies to optimize the transmit powers in the network. We demonstrate that employing sectorized antenna elements mitigates the detrimental effects of pilot contamination by rejecting a portion of interfering pilots in the spatial domain during channel estimation phase. Simulation results with practical sectorized antennas reveal that sectorization and multi-point coordination combined with sectorization lead to more than 1.7$\times$ and 2.6$\times$ improvements in the 95$\%$-likely per-user throughput, respectively. 
\end{abstract}


\section{Introduction} \label{sec:intro}
The development of new technologies and applications such as virtual reality, ultra high definition video conferencing, and internet of everything has caused a substantial increase in the demand for data rate in cellular systems \cite{morgado2018survey}. While conventional wireless networks are not able to meet this demand, massive  multi-input multi-output (MIMO) systems with time-division duplexing (TDD) provides a scalable solution \cite{larsson2014massive}. In massive MIMO cellular systems, a large number of antennas are employed at each base station (BS) to serve multiple single-antenna users within each cell over the same time-frequency interval.


Massive MIMO systems provide several advantages over conventional systems such as higher spectral and energy efficiency and simpler resource allocation and power optimization schemes \cite{larsson2014massive}. It is shown in \cite{marzetta2006much} that a single cell massive MIMO system operating in TDD has unbounded asymptotic capacity in the limit of an increasing number of antennas at the BS. More precisely, it is proved that the received signal power grows proportionally to number of antennas, whereas the interference plus noise power does not. References \cite{marzetta2010noncooperative} and \cite{jose2011pilot} demonstrate that under independent Rayleigh fading channel, there is a limit for the spectral efficiency in a multi-cell massive MIMO system due to a phenomenon called \textit{pilot contamination} which is caused by reusing pilot sequences used for uplink channel estimation in the TDD protocol. As the channel coefficients are time-varying, there is a limited time for channel estimation which leads to a finite number of orthogonal pilot sequences. Therefore, in order to support multiplicity of users within each cell in a large network, pilot sequences cannot be orthogonal across cells. 
As a result, the estimated channel vector of a user is contaminated by the non-orthogonal pilot sequences used by other users in the system. Pilot contamination has two detrimental effects: \textit{i}) it reduces the coherent array gain leading to a lower desired received signal power and \textit{ii}) it introduces a coherent interference whose power is proportional to the number of antennas at the BS under uncorrelated Rayleigh fading channel assumption. This coherent interference is the limiting factor for the asymptotic spectral efficiency in the limit of an infinite number of antennas at the BS \cite{marzetta2010noncooperative}.

Several methods have been proposed to alleviate the effects of pilot contamination in massive MIMO systems. It has been shown that smart assignment of pilot sequences to the users can reduce the effects of pilot contamination, such as assigning the same pilot sequence to users with different spatial channel features \cite{yin2015robust}. Also, reference \cite{tabu-search} presents a low-complexity iterative search-based method for pilot assignment to users so as to mitigate pilot contamination. An alternative approach is to reuse a pilot sequence less frequently in the system, i.e. using a larger pilot reuse factor as discussed in \cite{bjornson2016massive}. However, this method increases the number of time-frequency resource blocks spent on the channel estimation. In \cite{soft-pilot-reuse}, a soft pilot reuse scheme is investigated where the pilots assigned to the edge users are reused less frequently in the system so as to mitigate pilot contamination and increase the throughput for such users. Reference \cite{pow-opt-pilot} proposes a power optimization scheme based on the large-scale channel coefficients for the pilot transmission stage to mitigate the channel estimation error caused by the pilot contamination. 
Another technique is to use uplink pilot transmission jointly with uplink data transmission to improve the channel estimation quality \cite{ngo2012evd}, \cite{muller2014blind}. However, semi-blind channel estimation is required in this case as the uplink data is unknown at the BS \cite{yin2015robust,hu2016semi}. Multi-cell pilot contamination precoding is another technique which can be utilized to remove the coherent interference caused by pilot contamination \cite{ashikhmin2012,pcp1}. In this approach, an extra layer of precoding (for downlink) and decoding (for uplink) based on the large-scale fading coefficients are applied at each BS which require multi-cell coordination. Interference-rejecting precoding and decoding schemes such as multi-cell MMSE can be applied to remove the coherent interference caused by pilot contamination in the spatial domain \cite{bjornson2018massive,neumann2017}. However, it is usually difficult to provide a closed-form expression for the user spectral efficiency when such precoding and decoding schemes are applied \cite[Section 4.1]{bjornson2017massive} which in turn increases the complexity of power optimization. Furthermore, these precoding/decoding methods usually require estimation of inter-cell channel vectors (between each BS and the users in other cells) or the second order statistics of the channel vectors which escalates the complexity \cite[Section 4.1]{bjornson2017massive}. Reference \cite{bjornson2018massive} shows that for any fixed number of cells and under mild conditions on the users' channel statistics (without uncorrelated Rayleigh fading assumption), the asymptotic capacity of multi-cell massive MIMO systems is unlimited if multi-cell MMSE precoding/decoding is used at the BSs. 

Most of the literature on massive MIMO considers omnidirectional BS antennas with isotropic radiation patterns. However, in practice, the radiation pattern of antennas are not isotropic. For instance, microstrip antennas which are widely proposed in the literature \cite{larsson2014massive,gao2015massive,xingdong2014design,gao2016stacked,demoLund}, have relatively directional radiation patterns. \textcolor{black}{One of the contributions in this paper is to consider a sectorized radiation model to capture this directionality. In this model, the radiation pattern is characterized by a constant main-lobe gain over the angular interval of the beamwidth as well as a constant back-lobe gain elsewhere. We use this model to study the impact of three-fold sectorization in massive MIMO systems where the beamwidth is considered to be $120^\circ$ for each antenna element. The described sectorized radiation model, which has also been used in our recent work \cite{shahram2017}, is ideal due to the constant main-lobe and back-lobe gains as well as the sharp roll-off on the edge between the main-lobe and the back-lobe regions. Consequently, using this model leads to an upper-bound for the performance which can be used as a benchmark. To investigate practically achievable gains of sectorization, we also consider the radiation pattern of a microstrip antenna in our evaluations. This antenna also has a cone-shaped conductive reflector to flatten the directivity gain in the main-lobe region and reduce the back-lobe radiation.}

Multi-point coordination is envisioned to play an important role in fifth generation of wireless networks (5G) and beyond \cite{soret2018}. This coordination can be used to improve the performance in different ways such as mitigating inter-cell interference and providing macro-diversity or multiplexing gains. \textcolor{black}{In this paper, the objective is to study the impact of three-fold sectorization as well as multi-point coordination (with or without sectorized antennas) on the downlink performance of a massive MIMO system.} To this end, we consider two general communication settings: \textit{i}) \textit{sectorized setting} where each user is connected to one of the three antenna arrays located on the non-adjacent corners of the hexagonal cell that the user belongs to and \textit{ii}) \textit{Coordinated multi-point (CoMP) setting} where each user is served by coherent transmission of all three antenna arrays serving that cell. In the CoMP setting, three antenna arrays precode the downlink signals separately based on the estimated channel vectors and transmit them simultaneously. For both settings, we consider linear precoding schemes at each antenna array namely maximum-ratio (MR) and zero-forcing (ZF). \textcolor{black}{Note that optimizing precoding is computationally prohibitive since it has to be performed frequently. Consequently, lower complexity precoding schemes such as MR and ZF with optimized transmit powers have higher chances to be implemented in practice.} Using a standard information theoretic approach, we provide a closed-form lower-bound expression on the achievable downlink rate of the users as a function of large-scale fading coefficients for every combination of the described communication settings and precoding schemes. Using these expressions, we formulate the downlink power optimization problem with two fairness criteria: \textit{i}) network-wide max-min fairness (NMF) and \textit{ii}) per-cell max-min fairness (PMF). It should be noted that max-min fairness has been adopted in several other studies in the MIMO literature such as multi-user MIMO \cite{liu2013max} and cell-free massive MIMO \cite{ngo2017cellfree}. \textcolor{black}{However, in most of such studies, the power optimization problem is solved centrally, creating a large communication overhead.
Furthermore, a central solution requires high computational power.} 

Another contribution of this paper is to provide a wide range of power control strategies in terms of the required communication overhead and computational complexity. We first prove that the  optimization problem under NMF is quasi-concave and develop a centralized power control strategy using bisection method for sectorized and CoMP settings. One of the major issues with NMF is the computational complexity as the problem is solved for the entire network. Another downside of NMF is that a user with weak channel quality compromises the performance of all other users in the network due to the max-min fairness across the network. To resolve this issue, we next formulate power optimization under PMF for sectorized and CoMP settings in which we define max-min fairness separately within each cell. Furthermore, we provide closed-form solutions to the PMF problem for various communication settings which can be used by the central entity to solve the PMF problem very fast for the entire network. However, even with PMF, the amount of information that needs to be exchanged between the BSs and the central entity is large and is proportional to the number of the cells. To address this, we devise decentralized power control strategies in which the power optimization problem is solved locally at each BS. In this approach, each BS exchanges information with its direct neighboring BSs and solves the power optimization problem for its own users. We show that a combination of PMF and decentralized optimization provides a practical low-complexity and low-overhead solution for power optimization in sectorized and CoMP settings.    

\textcolor{black}{We use the developed analytical framework to evaluate the performance of various communication settings and power control strategies with extensive numerical examples providing several practical insights. A summary of the most important practical conclusions are: } 

\begin{itemize}[leftmargin=*]
\item \textcolor{black}{Using sectorized antennas can effectively alleviate pilot contamination by attenuating the interfering pilots received from the back-lobe of the antenna reception (radiation) pattern and enhance channel estimation in massive MIMO systems leading to substantial gains in the users' throughput. In effect, in terms of mitigating pilot contamination,  three-fold sectorization is somewhat similar to increasing pilot reuse factor from one to three except that sectorization does not impose any overhead while increasing pilot reuse factor require investing more resources on channel estimation.} 

\item \textcolor{black}{Multi-point coordination with omnidirectional antennas is not very appealing as it boosts the impact of pilot contamination fading away the macro-diversity gains of multi-point coordination. However, using multi-point coordination along with sectorized antennas at each BS leads to significant improvement as using sectorized antennas improves the channel estimation and multi-point coordination provide macro-diversity gains.}

\item \textcolor{black}{The proposed distributed power control strategy using PMF provides significant improvements compared to the case without power optimization in the $95\%-likely$ per-user throughput while having manageable computation complexity and low overhead.} 

\item \textcolor{black}{In a practical setting with the antenna radiation pattern of a practical patch antenna as well as  distributed power optimization with PMF, three-fold sectorization and multi-point coordination combined with three-fold sectorization lead to more than 1.7$\times$ and 2.6$\times$ improvements in the 95$\%$-likely per-user throughput, respectively, compared to the benchmark setting with omnidirectional antennas and without multi-point coordination.}  
\end{itemize}

\textit{Notation}: 
The set of numbers $\{1,2,\ldots, n\}$, $n\in \mathbb{N}$ is represented by $[n]$. Vectors and matrices are represented by lower-case and upper-case bold letters, respectively. Scalars are represented by lower-case non-bold letters. $\bff{I}_K$ and $\bff{0}_K$ are $K \times K$ identity and all-zero matrices, respectively. $[\bff{Y}]_{mk}$ is element $(m,k)$ of matrix $\bff{Y}$. 
The diagonal matrix of vector $\bff{v}$ is denoted by $\bff{D}_{\bff{v}}$. 
We use $P\textbackslash\{l\}$ to exclude member $l$ from set $P$. 
$\delta_{kk'}$ represents the discrete Dirac delta function. 

\section{System Model} \label{sec:sys-model}
\begin{figure}[h]
  \centering
  \includegraphics[width=0.38\linewidth, height=0.36\linewidth]{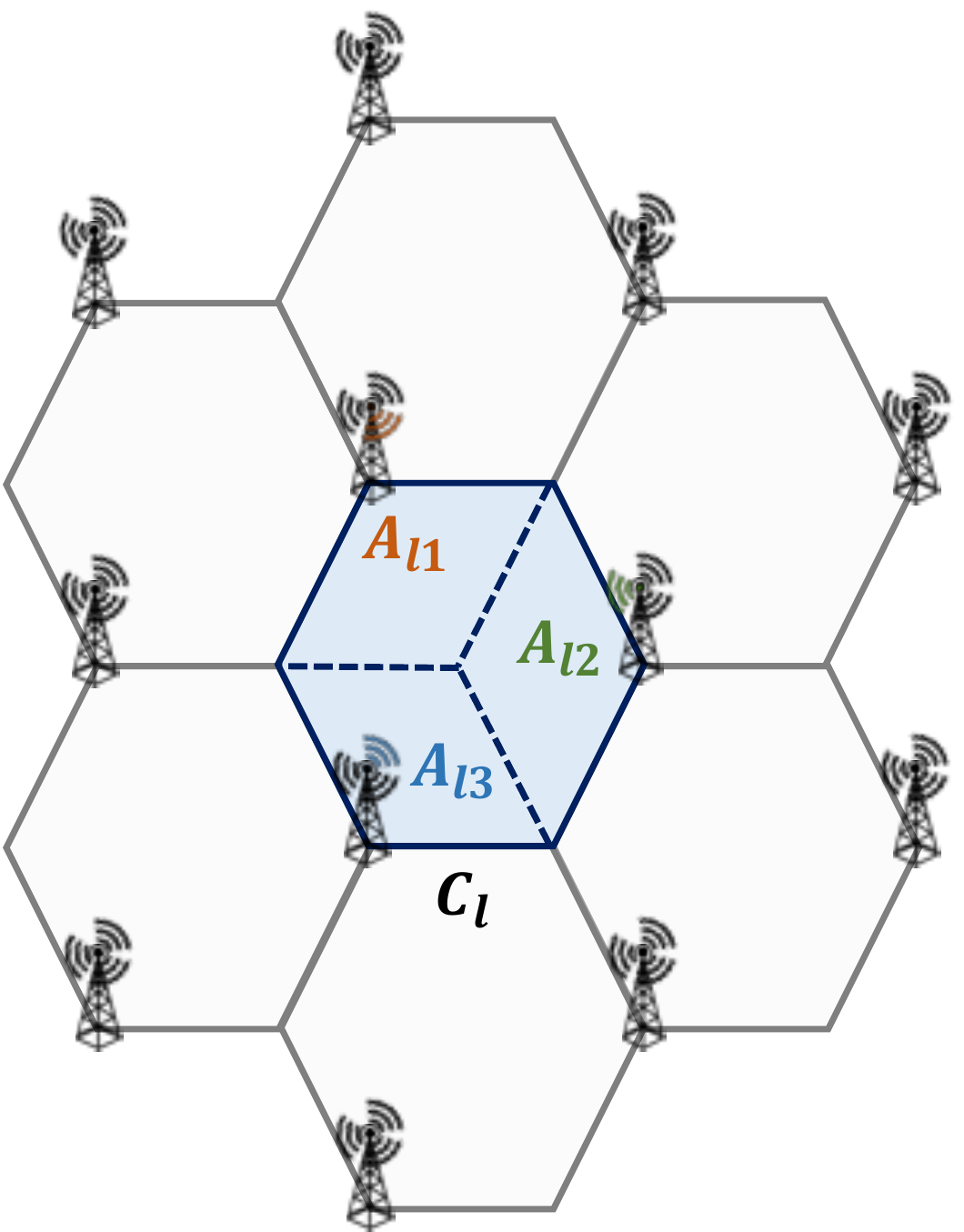}
  
  \caption{Multi-cell network with three-fold sectorization}
  \label{fig:sys-model}
  
\end{figure}

We consider the downlink of cellular system consisting of $L$ hexagonal cells each with $K$ users. 
Let $C_l, l\in[L]$ and $U_{lk}, k \in [K]$ denote cell $l$ and user $k$ in cell $l$, respectively. Three-fold sectorization is done such that three BSs are located at the non-adjacent corners of each cell, as shown in Figure \ref{fig:sys-model}. Each BS is composed of three antenna arrays such that each array serves one of the three neighboring cells. We assume that each antenna array has $M$ directional antenna elements whose orientations are aligned to cover the corresponding $120^{\circ}$ sector. As a result, there are $M_{\text{BS}} = 3M$ elements at each BS. 
Each antenna array is uniquely identified by a cell-sector pair $(l,i)$, and is denoted by $A_{li}$, where $i \in [3]$ is the sector index (see Figure \ref{fig:sys-model}). We also assume that user has a single omnidirectional antenna. \textcolor{black}{While we consider three-fold sectorization in this paper, it is straightforward to generalize the model, the analytical results, and the proposed algorithms for the case where each cell is divided into $3s, s\in\mathbb{N}$ sectors, each covered by an antenna array consisting of $M$ directional elements. We focus on the case with $s=1$ as it is more appealing from a practical perspective.}

\subsection{Antenna Directionality} \label{subsec:antenna}
\begin{figure}[h]
  \centering
  \includegraphics[width=0.65\linewidth]{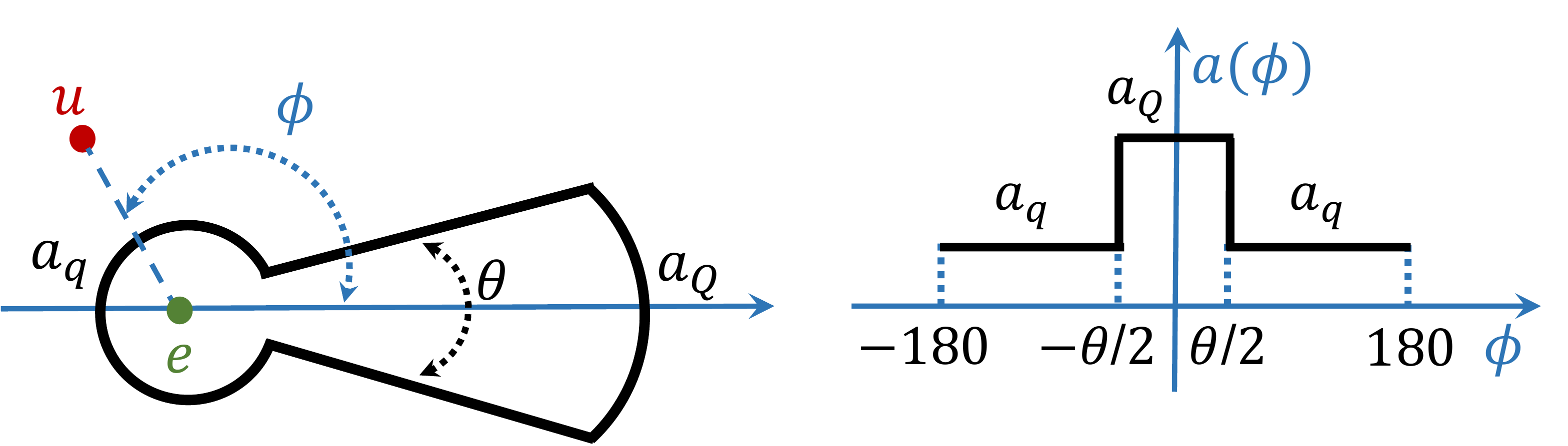}
  
  \caption{Ideal directional radiation pattern. $a(\phi)$ is the directivity gain in direction $\phi$. }
  \label{fig:pattern}
  
\end{figure}

\begin{figure*}
  
    \centering
        \begin{subfigure}[b]{0.22\textwidth}
            \includegraphics[width=\textwidth]{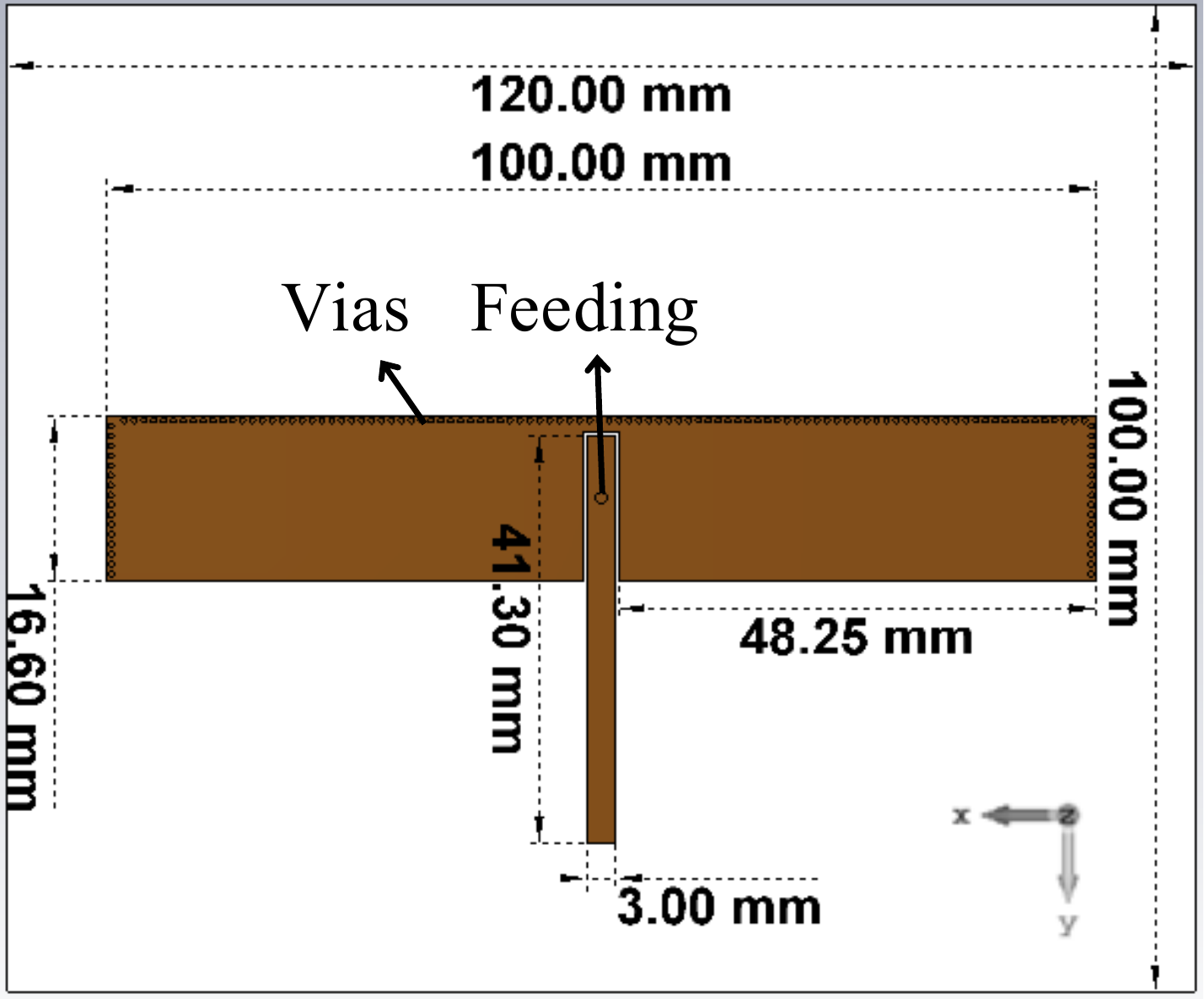}
            \caption[]%
            {{\small }}    
            \label{fig:patch-schematic}
        \end{subfigure}
        \begin{subfigure}[b]{0.3\textwidth}              \includegraphics[width=\textwidth]{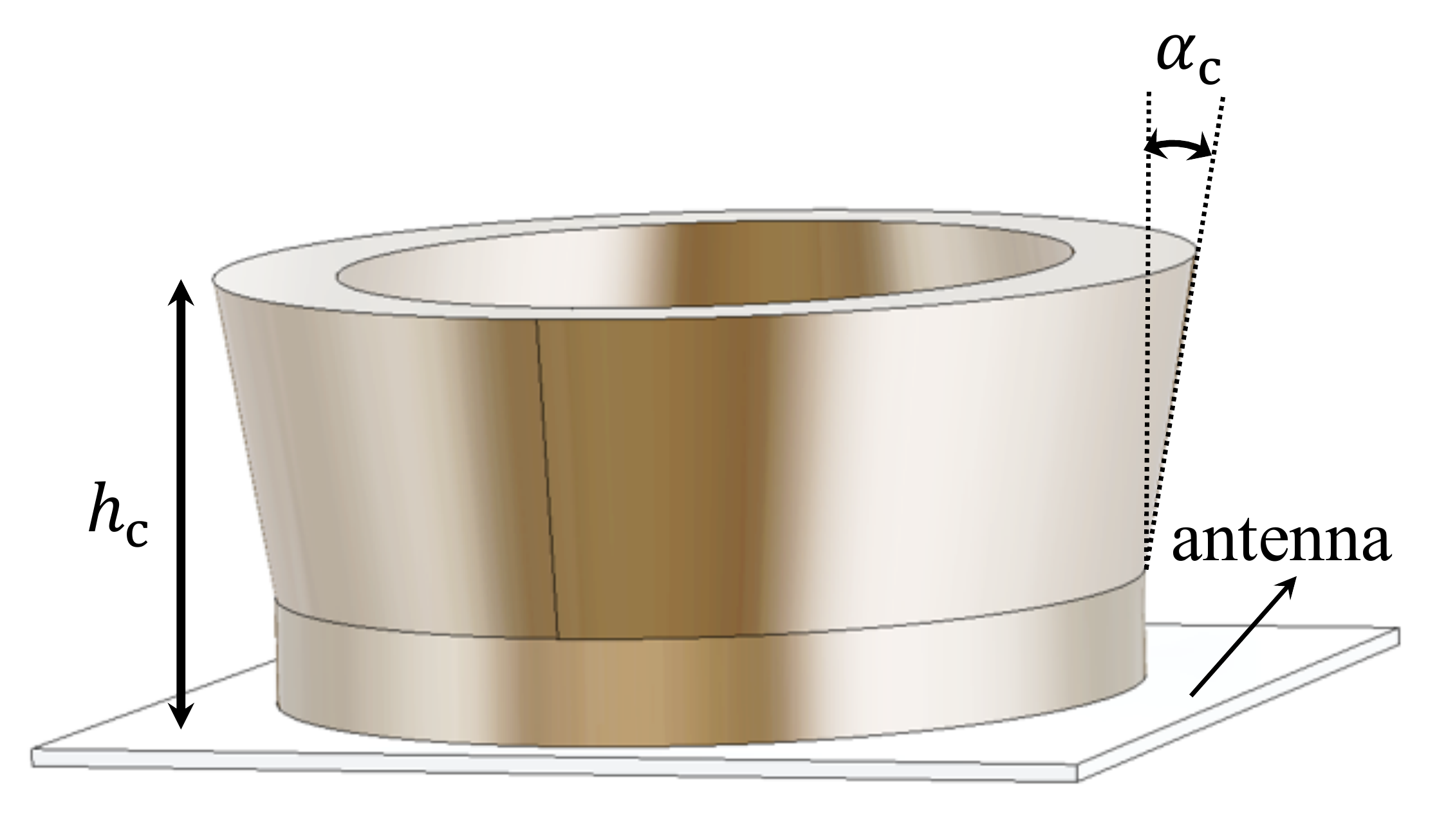}
            \caption[]%
            {{\small}}    
            \label{fig:cone}
        \end{subfigure}
        \begin{subfigure}[b]{0.4\textwidth}              \includegraphics[width=\textwidth]{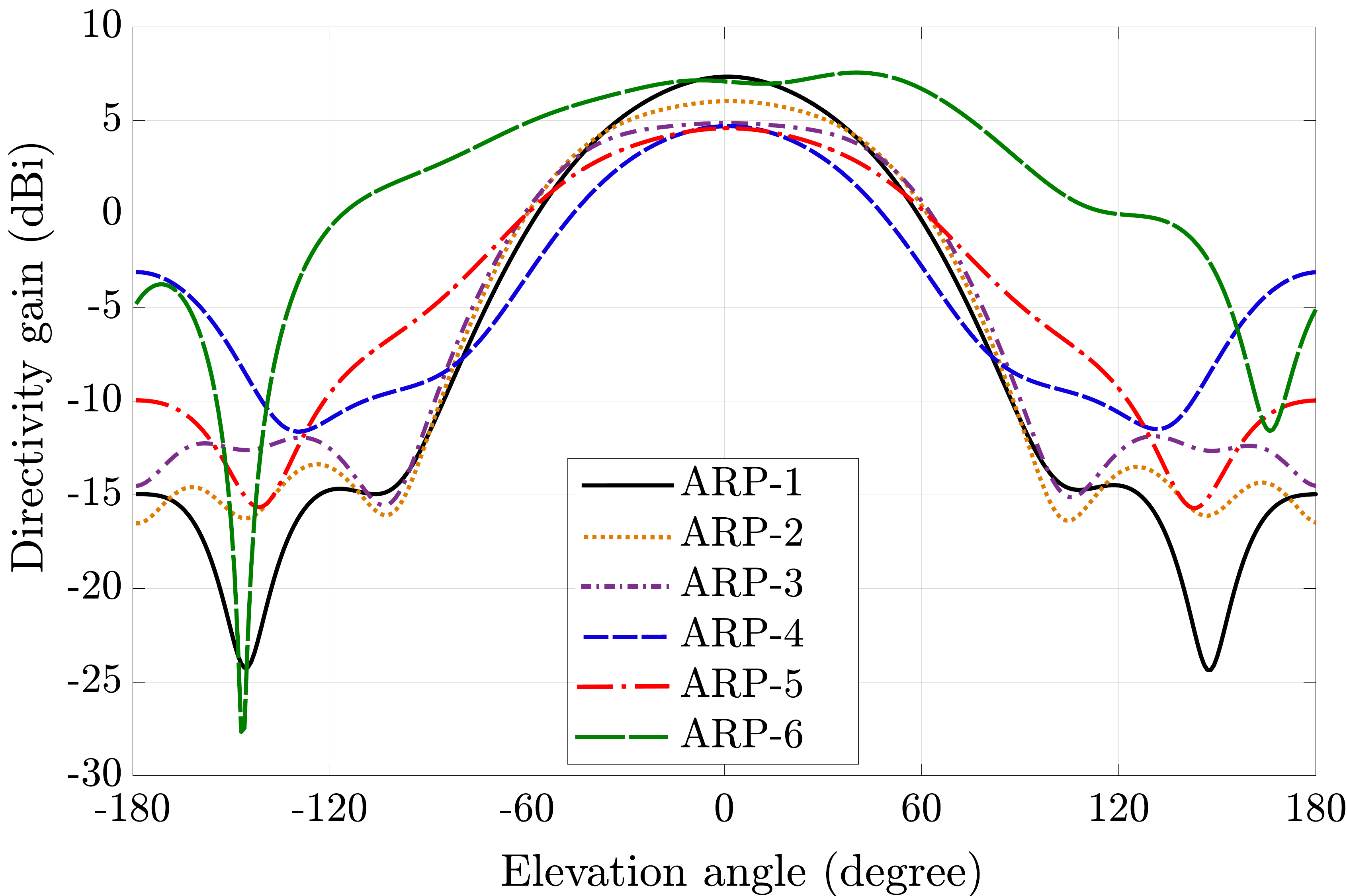}
            \caption[]%
            {{\small}}    
            \label{fig:patch-pattern}
        \end{subfigure}
        
        \caption{(a) Antenna element schematic, (b) conductive cone-shaped reflector placed on the antenna, (c) antenna radiation pattern for different design parameters listed in Table \ref{tab:arp}.}

\end{figure*}

Directional antenna elements radiate and receive most of the energy in certain directions. This effect is captured by a parameter called \textit{directivity gain} which determines the radiated (or received) power density in different angular directions around the antenna \cite[Chapter 2]{balanis}. 
To elaborate, if a single-antenna transmitter sends signal $x$, through a point to point channel $g$, the received signal is $y=\sqrt{a(\phi)}gx$, where $a(\phi)$ is the directivity gain which depends on the angular position of the receiver with respect to the transmitter as well as the type and shape of the antenna. We let $a^{li}_{l'k}$ denote the directivity gain between the antenna elements of array $A_{li}$ and user $U_{l'k}$. As the distance between the user and antenna array is substantially larger than the distance among the elements in the array, we assume that $a^{li}_{l'k}$ is independent of antenna index in the array and only depends on the angular position of the user. Let $\bff{a}^{li}_{l'} \triangleq [a_{l'1}^{li},\ldots, a_{l'K}^{li}]$ denote the \textit{directivity vector} between $A_{li}$ and the users in cell $C_{l'}$. 
In this paper we consider two different models for the antenna element radiation pattern:

\begin{enumerate}[wide=0pt]
\item \textit{Ideal Radiation Pattern (IRP)}: 
In this model, we adopt an ideal directional radiation pattern characterized by three parameters: \textit{i}) beamwidth $\theta$, \textit{ii}) main-lobe directivity gain $a_\text{Q}$ within the range of the beamwidth angular interval, and \textit{iii}) back-lobe directivity gain $a_\text{q}$ outside the beamwidth angular interval \cite{ramanathan2001performance}. We assume a lossless antenna model which implies that $\frac{\theta}{360}a_\text{Q}+(1-\frac{\theta}{360})a_\text{q}=1$, and $a_\text{Q} \leq \frac{360}{\theta}$ due to the conservation of power radiated in all directions~\cite{ramanathan2001performance}. Figure \ref{fig:pattern} depicts the described radiation pattern. This model will help to investigate the ultimate gains of three-fold sectorization by using $\theta=120^\circ$, $a_\text{Q}=3$, and $a_\text{q}=0$.

\item \textit{Actual Radiation Pattern (ARP)}: In this model, we consider the radiation pattern of a microstrip antenna element formed by placing two semi-dipole antennas perpendicular to each other as illustrated in Figure \ref{fig:patch-schematic}. The horizontal dipole is grounded with several vias and the vertical dipole is fed using a coaxial cable. In order to reduce back-lobe radiation and flatten the directivity in the main-lobe angular region, a conductive cone-shaped reflector is placed on the antenna as depicted by Figure \ref{fig:cone}. We can modify the radiation pattern parameters such as half-power beamwidth and flatness of the pattern in the beamwidth interval by changing the cone parameters such as height ($h_\text{c}$) and tapering angle ($\alpha_\text{c}$). Figure \ref{fig:patch-pattern} depicts the radiation pattern of this antenna element in the elevation plane for different cone parameters listed in Table \ref{tab:arp}. We observe that changing reflector parameters affects the beamwidth and back-lobe radiation. We can investigate practically achievable gains of sectorization by adopting ARP. 
\end{enumerate}

\begin{table}[H]
\centering
\caption{Different design parameters in ARP}

\scalebox{1}{
\begin{tabular}{cccc} \toprule
$\bf Pattern~index$ & $\bf Cone~height$ ~$\boldsymbol{h_\text{c}}$    & $\bf Tapering~angle$ ~$\boldsymbol{\alpha_\text{c}}$ & $\bf 3~dB~beamwidth$         \\ \midrule
ARP-1&30 mm  &$0^\circ$ & $74^\circ$\\
ARP-2&30 mm  &$15^\circ$ & $96^\circ$\\
ARP-3&30 mm  &$20^\circ$ &$110^\circ$\\
ARP-4&5  mm  &$20^\circ$ &$75^\circ$\\
ARP-5&15 mm  &$20^\circ$& $100^\circ$\\
ARP-6 (without reflector)&0 mm  &$0^\circ$ & $151^\circ$\\
\bottomrule
\end{tabular}
}
\label{tab:arp}

\end{table}

\subsection{Channel Model}
Orthogonal Frequency-Division Multiplexing (OFDM) is a prominent technique which utilizes discrete Fourier transform to decompose the channel into parallel channels called \textit{subcarriers} in order to compensate for frequency selectivity. For a given subcarrier in the system, we define the propagation channel coefficient between antenna element $m\in[M]$ of the array $A_{li}$ and user $U_{l'k}$ as $g_{l'k}^{mli} = \sqrt {\beta_{l'k}^{li}}, h_{l'k}^{mli}$
where $\beta_{l'k}^{li}$ is the large-scale fading coefficient, which depends on the shadowing and range dependent path-loss between the corresponding user and the antenna element, and $h_{l'k}^{mli}\in \mathbb{C}$ is the small-scale fading coefficient modeling multi-path fading. As the
distance between a user and an antenna array is substantially larger than the distance among the elements in the antenna array, we assume that the large-scale fading coefficients are independent of the antenna element index $m$. \textcolor{black}{We adopt uncorrelated Rayleigh fading model in which $h_{l'k}^{mli}, \forall(m,l,i,l',k)$ are \textit{i.i.d.} random variables with complex normal distribution $\mathcal{CN}(0,1)$. This model has been used along with sectorized antennas in several prior works in the literature of MIMO systems such as \cite{sun2020interference, huang2009increasing}. Additionally, in this paper we consider the regime with moderately large number of antennas ($\sim100$ antennas) at the BS antenna arrays for the sake of practicality. It can be seen from the results reported in \cite[Chapter 4]{bjornson2017massive} that in such a regime, the result of spatially correlated fading model is sufficiently close to those of uncorrelated fading model especially for MR and ZF precoding schemes which are the choices in our paper. Consequently, it is expected that our numerical results mimic the ones with correlated channel model closely.} We assume that the small-scale and large-scale fading coefficients are constant over small-scale and large-scale coherence blocks represented by $T_\text{c}$ and $T_\beta$, respectively. We further assume that the channel response is constant over a range of frequencies called coherence bandwidth denoted by $B_\text{c}$. 
As a result, the channel coefficients are assumed to be static in a time-frequency segment called \textit{coherence interval}, with time duration $T_\text{c}$ and frequency duration $B_\text{c}$ \cite[Chapter 2]{marzetta2016fundamentals}. We denote the number of complex samples within each coherence interval by $\tau_\text{c}=T_\text{c}B_\text{c}$. Furthermore, we neglect the overhead due to the cyclic prefix. 
To facilitate the derivations, we will use $\gbf_{l'k}^{li} \in \mathbb{C}^{M\times1}$
to denote {\em channel vector} between $A_{li}$ and $U_{l'k}$ and $\bff{G}^{li}_{l'} \triangleq [\gbf_{l'1}^{li},\gbf_{l'2}^{li},\ldots, \gbf_{l'K}^{li}]\in \mathbb{C}^{M\times K}$ to denote the \textit{channel matrix} between $A_{li}$ and the users in cell $C_{l'}$. 


\subsection{Communication Settings} \label{subsec:comm-setting}
Generally, a user can be served by several nearby BSs. We assume that user $U_{lk}$ is served by either one or all three antenna arrays serving cell $C_l$. Let $e_{lik}$ denote the binary variable indicating the user to antenna array association which is one if $U_{lk}$ is associated with $A_{li}$ and zero otherwise.
Let $K_{li}$ denote the number of the users associated with antenna array $A_{li}$, i.e. $K_{li}=\sum_{k \in [K]}e_{lik}$. Moreover, let $\mathcal{A}_{li}$ be the set of the user indices associated with antenna array $A_{li}$ that is $\mathcal{A}_{li}=\{k: e_{lik}=1\}$, $(l,i)\in [L]\times[3]$. We consider three different communication settings based on user to antenna array association described as follows.
\begin{enumerate}[wide=0pt]
\item \textit{Sectorized setting with Minimum Distance criterion (SecMD):}
In this setting, user $U_{lk}$ is associated with the closest antenna arrays among $A_{l1}$, $A_{l2}$, and $A_{l3}$.

\item \textit{Sectorized setting with Maximum average received Power criterion (SecMP):}
In this setting, user $U_{lk}$ is associated with the antenna array from which it receives the maximum average power. 
We note that SecMP takes into account the effect of antenna directivity, path loss and shadowing while SecMD only considers the path loss which is inversely proportional to the distance.   
\item \textit{Coordinated Multi-point setting with sectorized antennas (CoMPSec):}
In this setting, the users in each cell are served by all the three antenna arrays serving cell $C_l$, i.e. $e_{lik}=1$, $i\in[3]$. As the user equipment has a single antenna, being served by multiple antenna arrays can only lead to \textit{diversity} and \textit{power gains}. In this case, all the antenna arrays serving a cell should have the downlink information intended for the users within that cell.    
\textcolor{black}{
\item \textit{Coordinated Multi-point setting with omnidirectional antennas (CoMPOmn):} To highlight the importance of sectorization when having CoMP, we also consider the setting where CoMP is used without sectorization where each BS has only one array consisting of $3M$ omnidirectional antennas. The BSs are still located on the non-adjacent corners of the cells and each cell is served by three BSs. While we do not provide the analytical results for this case due to the space limit and similarity to Setting 3, we have provided the simulation results in Section \ref{sec:sim}.}
\end{enumerate}

\subsection{Time-Division Duplexing (TDD) Protocol} \label{subsec:tdd-protocol}
\begin{figure}[t]
  \centering
  \includegraphics[width=0.50\linewidth]{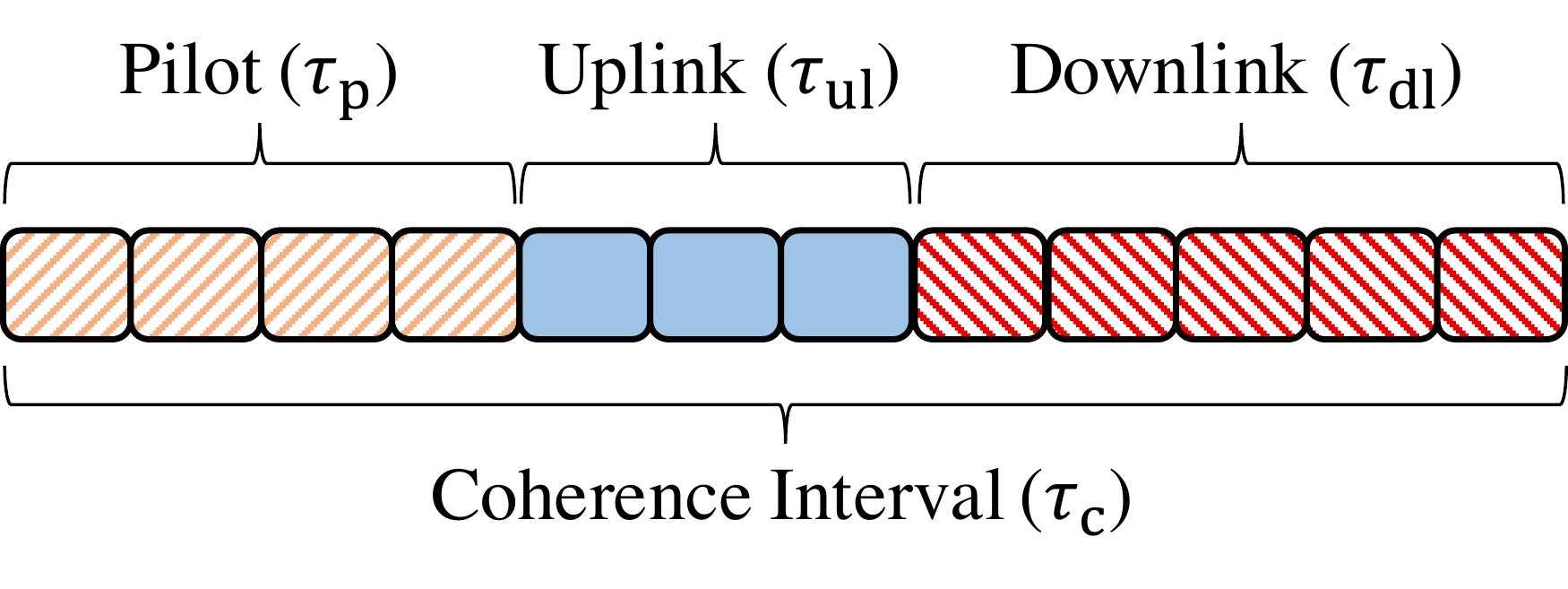}
  
  \caption{Configuration of time-frequency samples in a coherence interval}
  \label{fig:coh-interval}
  
\end{figure}

Channel state information (CSI) is essential to fully exploit the multiplexing gains in massive MIMO systems. It is known that using a TDD protocol with uplink channel estimation provides a scalable system where we can increase the number of antennas at the BS without increasing the channel estimation overhead \cite{larsson2014massive}.
In this paper, we consider the TDD protocol where the time-frequency samples at each coherence interval are used for uplink channel estimation, uplink data transmission, and downlink data transmission as illustrated by Figure \ref{fig:coh-interval}. The number of samples assigned to channel estimation, uplink data, and downlink data are denoted by $\tau_{\text{p}}$, $\tau_{\text{ul}}$, and $\tau_{\text{dl}}$, respectively. Therefore, we have $\tau_{\text{c}}=\tau_{\text{p}}+\tau_{\text{ul}}+\tau_{\text{dl}}$. We neglect the overhead due to cyclic prefix or processing but it is straightforward to include them as they would not change the analyses.

We use the standard TDD protocol as in \cite{marzetta2016fundamentals,pcp1} which is described briefly subsequently.
In the beginning of each large-scale coherence block, each BS estimates the large-scale fading coefficients between itself and all the users in the network. If a centralized power control scheme is used, large-scale channel coefficients are sent to a central entity in the network for power optimization. Otherwise, large-scale channel coefficients are exchanged locally to perform power optimization. Next, each antenna array sends a \textit{long-term effective channel gain} to every user connected to that array. These gains are functions of large-scale channel coefficients, power coefficients, and the applied precoding scheme. Users use this information to decode their downlink signal. Next, all users transmit their pilot sequences followed by their uplink signals. After that, each antenna array estimates the channel vector between itself and the users located in its cell by processing the received pilot sequences. Then, each array decodes the received uplink signals using the estimated channel vectors. Next, antenna arrays precode the downlink signals using the estimated channel vectors. All arrays synchronously transmit the prepared signals. Eventually, each user decodes its downlink signal using the long-term effective channel gains received from the antenna arrays serving that user. 

For the TDD protocol given above, we assume that each array $A_{li}$ can accurately estimate and track the combined quantities $a^{li}_{l'k}\beta^{li}_{l'k},(l',k)\in[L]\times [K]$ required for power optimization discussed in Section \ref{sec:power-opt}. Note that these quantities do not depend on the antenna element index and vary slowly in time. Consequently, it is possible to estimate them through a similar process as the one introduced in \cite[Section III.D]{pcp1-v1} for estimation of large-scale channel coefficients with omnidirectional antennas. We also assume that $A_{li}$ has the means to forward long-term effective channel gains to the users in $C_l$. As in \cite{pcp1}, we will not consider the resources required for implementing these assumptions. Also, we assume that the channel coefficients and antenna directivity gains are similar for uplink and downlink due to the reciprocity provided in TDD. 

\textcolor{black}{In practice, several factors such as hardware mismatch between transmitter and receiver can violate the reciprocity. Since reciprocity is essential in the operation of TDD massive MIMO systems, several prior studies such as 
\cite{calibration2} have investigated the factors violating reciprocity and proposed calibration techniques to compensate them.}

\section{Channel Estimation} \label{sec:chann-est}
According to the TDD protocol described in Section \ref{subsec:tdd-protocol}, the intra-cell channel vectors are estimated by the antenna arrays using the received uplink pilot sequences. 
Let $\bff{\Phi}_l \in \mathbb{C}^{\tau_\text{p} \times K}$ denote the pilot matrix which includes $K$ orthonormal pilots used in cell $C_l$. We assume that these pilots are assigned randomly to the users. Let $\boldsymbol{\phi}_{lk}\in \mathbb{C}^{\tau_\text{p} \times 1}$ be the pilot vector assigned to user $U_{lk}$. As reusing a pilot sequence within the same cell can lead to severe pilot contamination, we assume that the pilot sequences are orthonormal in each cell 
which implies that $\boldsymbol\phi_{lk}^H \boldsymbol\phi_{lk'}=\delta_{kk'}$ for any $k,k' \in [K]$ and $l\in[L]$. Therefore, collectively we have $\bff{\Phi}^H_l\bff{\Phi}_l=\bff{I}_K$. 
Note that since the number of orthogonal $\tau_\text{p}$-tuples can not exceed $\tau_\text{p}$, we have $K\leq\tau_\text{p}$, i.e. at most $\tau_\text{p}$ users can be served at each cell while using orthonormal pilots within each cell \cite{pcp1}.
Due to the independence of channel coefficients across different coherence intervals, channel estimation is repeated for each coherence interval $\tau_\text{c}$, hence $\tau_\text{p} \leq \tau_\text{c}$. Consequently, it is required to reuse the pilots in different cells. To facilitate the derivations, we assume that for any two distinct cells, the pilot sequences are either orthogonal from cell to cell, or they are replicated. Let $P_l$ denote the set of the cells in which we reuse the pilot sequences of the cell $C_l$. As a result we have $\bff{\Phi}^H_l\bff{\Phi}_{l'}=\bff{I}_K$ if $l'\in P_l$ and $\bff{\Phi}^H_l\bff{\Phi}_{l'}=\bff{0}_K$, otherwise.
Users of cell $C_l$ transmit $\bff{X}_{\text{p}l}=\sqrt{\tau_{\text{p}}}\bff{\Phi}_l^H$, $l \in [L]$ collectively during channel estimation phase. Array $A_{li}$ receives 
\begin{align}
\bff{Y}_{\text{p}li}&=\sum_{l'\in [L]} \sqrt{\rho_{\text{ul}}}\bff{G}^{li}_{l'}\bff{D}^{1/2}_{\bff{a}_{l'}^{li}}\bff{X}_{\text{p}l'}+ \bff{W}_{\text{p}li},
%
\end{align}
where $\rho_{\text{ul}}$ denotes the uplink signal to noise ratio (SNR) and the entries of the $M\times \tau_{\text{p}}$ receiver noise matrix, $\bff{W}_{\text{p}li}$, are \textit{i.i.d.} $\mathcal{CN}(0,1)$. Antenna array $A_{li}$ performs a de-spreading operation by correlating the received signals with each of the $K$ pilot sequences used in cell $C_l$. 
Next, $A_{li}$ performs a minimum mean square error estimation \cite{kay-estimation} to estimate $g^{lim}_{l'k}$ as follows.
\begin{align}
\hat{g}^{lim}_{l'k}=\frac{\sqrt{\rho_{\text{ul}}\tau_{\text{p}} a^{li}_{l'k}}\beta^{li}_{l'k}}{1+\rho_{\text{ul}}\tau_{\text{p}} \sum_{l''\in P_l} a^{li}_{l''k}\beta^{li}_{l''k}}y'_{mk}, \qquad l'\in P_l, \label{eq:channel-estimate-single2}
\end{align}
where, $y'_{mk}\triangleq[\bff{Y}_{\text{p}li} \bff{\Phi}_l]_{mk}=\sqrt{\rho_\text{ul}\tau_\text{p}}\sum_{l' \in P_l} \left(\sqrt{a^{li}_{l'k}}g^{lim}_{l'k}\right)+w'_{mk}$, $w'_{mk}\sim \mathcal{CN}(0,1)$.  The mean-square of the channel estimate $\hat{g}^{lim}_{l'k}$ is given by
\begin{align}
\gamma_{l'k}^{li} \triangleq \E{|\hat{g}^{lim}_{l'k}|^2}=\frac{\rho_{\text{ul}}\tau_{\text{p}}a^{li}_{l'k}\beta^{li}_{l'k}}{1+\rho_{\text{\text{ul}}}\tau_{\text{p}}  a^{li}_{l'k}\beta^{li}_{l'k}+\rho_{\text{\text{ul}}}\tau_{\text{p}} \sum_{l''\in P_l\textbackslash\{l'\}} a^{li}_{l''k}\beta^{li}_{l''k}}\beta^{li}_{l'k},  \qquad l'\in P_l. \label{eq:channel-estimate-single}
\end{align}
As a result of the pilot reuse, the received pilot of user $U_{l'k}$ is contaminated by the pilots received from users $U_{l''k}$, $l''\in P_l\textbackslash\{l'\}$. This phenomenon is called \textit{pilot contamination} which gives rise to the last term in the denominator of \eqref{eq:channel-estimate-single}. 
\textcolor{black}{According to \eqref{eq:channel-estimate-single}, we have $\gamma_{l'k}^{li}<\beta^{li}_{l'k}$, implying that the expected value of estimated channel power is always lower than its actual value due to the estimation error.} 
The channel estimation error is denoted by $\tilde{g}^{lim}_{l'k}\triangleq\hat{g}^{lim}_{l'k}-g^{lim}_{l'k}$, and the mean-square error is $\E{|\tilde{g}^{lim}_{l'k}|^2}= \beta^{li}_{l'k}-\gamma_{l'k}^{li}$. \textcolor{black}{Equation \eqref{eq:channel-estimate-single} also reveals the superiority of using sectorized antennas in massive MIMO systems. To elaborate, assuming a uniform spatial distribution for the interfering pilots, the numerator and one third of the interference terms (i.e. pilot contamination terms) in the denominator of $\gamma_{lk}^{li}$ are boosted by the antenna main-lobe gain while other two thirds of the interference terms are attenuated by the antenna back-lobe gain. Consequently, the impact of pilot contamination is reduced and the channel estimation quality is enhanced when using sectorized antennas, compared to the omnidirectional case where the directivity gains are all unity.} 
In matrix form, we can represent the estimate \eqref{eq:channel-estimate-single2} as $\hat{\bff{G}}^{li}_{l'}=\bff{Z}^{li}\bff{D}^{1/2}_{\boldsymbol{\gamma}_{l'}^{li}}$,
where $\hat{\bff{G}}^{li}_{l'}\in \mathbb{C}^{M\times K}$ is the matrix of channel estimates, whose $(m,k)$-th element is $\hat{g}^{lim}_{l'k}$. Furthermore, $\boldsymbol \gamma_{l'}^{li}=[\gamma_{l'1}^{li},\gamma_{l'2}^{li},\ldots,\gamma_{l'K}^{li}]^T$, and the elements of $\bff{Z}^{li}$ are \textit{i.i.d.} $\mathcal{CN}(0,1)$. We note that $\bff{Z}^{li}$ is independent of $l'$ which is an important property used in the performance analysis in Section \ref{sec:analysis}. 
Moreover, we define $\tilde{\bff{Z}}^{li}\in \mathbb{C}^{M\times K_{li}}$ as the submatrix of $\bff{Z}^{li}$ which is constructed by removing columns $k\not \in \mathcal{A}_{li}$ from $\bff{Z}^{li}$. This matrix will be used in the precoding stage in the next section.

\section{Performance Analysis} \label{sec:analysis}

In this section, we analyze the downlink performance of a massive MIMO system operating based on the TDD protocol described in Section \ref{subsec:tdd-protocol}. The analysis is performed in terms of the binary association variables $e_{lik}$, which are defined differently for sectorized and CoMP settings in Section \ref{subsec:comm-setting}. We consider two linear precoding schemes namely \textit{zero-forcing (ZF)} and \textit{maximum-ratio (MR)} and the objective is to derive closed-form lower-bound expressions on the per-user achievable rates for these precoding schemes.

Let $q_{lk} \in \mathbb{C}$, $(l,k) \in [L]\times[K]$ denote the desired downlink symbol of user $U_{lk}$. We assume that symbols $q_{lk},(l,k) \in [L]\times[K]$ are uncorrelated random variables with zero mean and unit variance. Let $\tilde{\bff{q}}_{li}\in \mathbb{C}^{K_{li}\times 1}$ be the vector consisting of the downlink symbols of the users connected to antenna array $A_{li}$. Moreover, let $\eta_{lik} \in [0,1]$ denote the power coefficient assigned to user $U_{lk}$ by antenna array $A_{li}$. We define $\tilde{\boldsymbol{\eta}}_{li} \in \mathbb{R}_{\geq0}^{K_{li}\times 1}$ as the vector of the power coefficients assigned by antenna array $A_{li}$ to the users connected to that array.
The antenna array $A_{li}$ precodes downlink symbol vector $\tilde{\bff{q}}_{li}$ as follows,
\begin{align}
    \bff{x}_{li} =\bff{B}_{li}\bff{D}^{1/2}_{\tilde{\boldsymbol{\eta}}_{li}} \tilde{\bff{q}}_{li} = \sum_{k\in [K]} \sqrt{\eta_{lik}}\bff{b}_{lik}e_{lik}q_{lk}, \label{eq:precoding}
\end{align}
where, $\bff{b}_{lik}$ is the $k$th column of the precoding matrix $\bff{B}_{li} \in \mathbb{C}^{M\times K_{li}}$. 
We assume that the power budget of each antenna array is a third of the total power budget of the BS, i.e. 
\begin{align}
  \E{||\bff{x}_{li}||^2} \leq 1/3,\;\;  (l,i)\in [L]\times[3] \label{eq:power-budget}. 
\end{align}

Furthermore, we assume that the precoding matrix $\bff{B}_{li}$ is normalized such that
\begin{align}
  \E{||\bff{x}_{li}||^2} =  \sum_{k\in\mathcal{A}_{li}} \eta_{lik}, \;\;(l,i)\in [L]\times[3]. \label{eq:power-scale}
\end{align}
Consequently, the power budget constraint \eqref{eq:power-budget} is equivalent to   
\begin{align} 
\begin{cases}
    \eta_{lik} = 0 ~~\text{if}~~ e_{lik}=0, \quad (l,i,k)\in[L]\times[3]\times[K], \\
    \sum_{k\in[K]} \eta_{lik}\leq \frac{1}{3}, ~~~~~~\quad (l,i)\in[L]\times[3] ,
\end{cases} \label{eq:power-coeff}
\end{align}
The first constraint in \eqref{eq:power-coeff} implies that each antenna array allocates power only to the users connected to that array and the second one ensures the sum-power constraint for the array. Next, antenna array $A_{li}$ transmits the precoded vector $\bff{x}_{li}$. Let $\bff{y}_l \in \mathbb{C}^{K\times 1}$ denote the vector received by $K$ users in cell $C_l$. We have
\begin{align}
\bff{y}_l&=\sqrt{\rho_{\text{dl}}}\sum_{l'\in [L]} \sum_{i\in[3]}\bff{D}^{1/2}_{\bff{a}^{l'i}_l}\bff{G}_{l}^{l'i^T}\bff{x}_{l'i}+\bff{w}_l \label{eq:yd-1}
\end{align}
where, $\rho_{\text{dl}}$ denotes the downlink SNR (i.e. maximum BS transmit power divided by noise power) and $\bff{w}_l \sim \mathcal{CN}(0,\bff{I}_K)$ is the downlink noise vector. Consequently, user $U_{lk}$ receives
\begin{align}
y_{lk}&=\underbrace{\sqrt{\rho_{\text{dl}}}\sum_{l'\in P_l} \sum_{i\in[3]}e_{l'ik}\sqrt{a^{l'i}_{lk}}\hat{\bff{g}}_{lk}^{{l'i}^T}\bff{x}_{l'i}}_{T_1}  +   \underbrace{\sqrt{\rho_{\text{dl}}}\sum_{l'\in P_l} \sum_{i\in[3]}e_{l'ik}\sqrt{a^{l'i}_{lk}}\tilde{\bff{g}}_{lk}^{{l'i}^T}\bff{x}_{l'i}}_{T_2} \notag\\
&+\underbrace{\sqrt{\rho_{\text{dl}}}\sum_{l'\in P_l} \sum_{i\in[3]}(1-e_{l'ik})\sqrt{a^{l'i}_{lk}}\bff{g}_{lk}^{{l'i}^T}\bff{x}_{l'i}}_{T_3}   +   \underbrace{\sqrt{\rho_{\text{dl}}}\sum_{l'\not\in P_l} \sum_{i\in[3]}\sqrt{a^{l'i}_{lk}}\bff{g}_{lk}^{{l'i}^T}\bff{x}_{l'i}}_{T_4} + \underbrace{w_{lk}}_{T_5}. \label{eq:yd-1k}
\end{align}
The term $T_1$ contains
the signal of interest for user $U_{lk}$ as well as the interference containing the signal of users $U_{l'k}$, $l'\in P_l\textbackslash\{l\}$. The remaining terms, i.e. $T_2$, $T_3$, $T_4$, and $T_5$, contribute to the effective noise. 
The variance of each of these four effective noise terms is independent of whether ZF or MR precoding is used. It can be shown that
\begin{align}
  &\Var{T_2}=\rho_{\text{dl}}\sum_{l'\in P_l} \sum_{i\in[3]} e_{l'ik} a^{l'i}_{lk}\left(\beta^{l'i}_{lk}-\gamma^{l'i}_{lk}\right)\left(\sum_{k'\in \mathcal{A}_{l'i}}\eta_{l'ik'}\right), 
  &&\Var{T_4}=\rho_{\text{dl}}\sum_{l'\not\in P_l} \sum_{i\in[3]}  a^{l'i}_{lk}\beta^{l'i}_{lk}\left(\sum_{k'\in \mathcal{A}_{l'i}}\eta_{l'ik'}\right),\notag\\
  &\Var{T_3}=\rho_{\text{dl}}\sum_{l'\in P_l} \sum_{i\in[3]} (1-e_{l'ik}) a^{l'i}_{lk}\beta^{l'i}_{lk}\left(\sum_{k'\in \mathcal{A}_{l'i}}\eta_{l'ik'}\right),
  &&\Var{T_5}=1\label{eq:T2-T5}.
\end{align}
Next, we decompose the term $T_1$ to separate the desired signal and interference. As the details of $T_1$ depends on the precoding scheme, we continue the analysis for particular choices of precoding matrix $\bff{B}_{li}$ modeling MR and ZF processing schemes.

\subsection{Zero-Forcing precoding} \label{subsec:zf}
In this case, the antenna array $A_{li}$ applies the precoding matrix $\bff{B}^{\text{zf}}_{li}=\sqrt{M-K_{li}}\tilde{\bff{Z}}^{li^*}\Big(\tilde{\bff{Z}}^{li^T}\tilde{\bff{Z}}^{li^*}\Big)^{-1}$. By employing the principles of random matrix theory, it can be shown that $\E{\bigg(\Big(\tilde{\bff{Z}}^{li^T}\tilde{\bff{Z}}^{li^*}\Big)^{-1}\bigg)_{kk}}=(M-K_{li})^{-1}$, $(l,i)\in [L]\times[3]$  \cite[Appendix A]{marzetta2016fundamentals}. Using this identity, it can be shown that \eqref{eq:power-scale} holds. 
Furthermore, by substituting $\bff{B}^{\text{zf}}_{li}$ in \eqref{eq:yd-1k} we have
\begin{align}
    T_1&=\underbrace{\sqrt{\rho_{\text{dl}}}q_{lk}\sum_{i\in[3]}e_{lik}\sqrt{(M-K_{li})a_{lk}^{li}\gamma_{lk}^{li}\eta_{lik}}}_{T^{\text{zf}}_{11}}+\underbrace{\sqrt{\rho_{\text{dl}}}\sum_{l'\in P_l\textbackslash\{l\}}q_{l'k}\sum_{i\in[3]}e_{l'ik}\sqrt{(M-K_{l'i})a_{lk}^{l'i}\gamma_{lk}^{l'i}\eta_{l'ik}}}_{T^{\text{zf}}_{12}} \label{eq:yd-zf-user},
\end{align}
where $T^{\text{zf}}_{11}$ is the desired signal of user $U_{lk}$ and $T^{\text{zf}}_{12}$ is the coherent interference due to reusing pilots in different cells. 
The coefficient of the desired signal in the received signal is called the \textit{effective channel gain}. According to \eqref{eq:yd-zf-user}, the effective channel gain for user $U_{lk}$ is $g^{\text{eff,zf}}_{lk}=\sum_{i\in [3]} g^{\text{eff,zf}}_{lik}$ when ZF is applied,
where $g^{\text{eff,zf}}_{lik}=e_{lik}\sqrt{(M-K_{li})\rho_{\text{dl}}a_{lk}^{li}\gamma_{lk}^{li}\eta_{lik}}$. We note that there is only one coefficient for user $U_{lk}$ which is the summation of three quantities (i.e. $g^{\text{eff,zf}}_{lik}$, $i\in[3]$) each of which corresponds to one of the antenna arrays serving cell $C_l$. Antenna array $A_{li}$ sends $g^{\text{eff,zf}}_{lik}$ to user $U_{lk}$ if ZF precoding is applied. User $U_{lk}$ then calculates $g^{\text{eff,zf}}_{lk}$ and use it for decoding. We note that $g^{\text{eff,zf}}_{lik}$ is a function of $K_{li}$, power coefficient $\eta_{lik}$ and  $a^{li}_{lk}\gamma^{li}_{lk}$ which is in turn a function of the quantities $a^{li}_{l'k}\beta^{li}_{l'k},l'\in P_l$ according to \eqref{eq:channel-estimate-single}. In Section \ref{sec:power-opt}, we will show that the optimized power coefficients are constant during each large-scale coherence block. Hence, $g^{\text{eff,zf}}_{lik}$ is sent to the user once at the beginning of each large-scale coherence block. The following theorem provides a lower-bound expression on the achievable downlink data rate when ZF precoding is applied at each antenna array.  
\begin{Theorem}\label{thm:zf}
For the multi-cell massive MIMO system with sectorized antennas and zero-forcing precoding at each antenna array, the downlink transmission rate of user $U_{lk}, (l,k)\in [K]\times[L]$, $R^{\text{zf}}_{lk}$, is lower bounded by
\begin{align*}
&R^{\text{zf}}_{lk} \geq (\tau_{\text{dl}}/\tau_{\text{c}})W\log_2\left(1+\text{SINR}^{\text{zf}}_{lk}\right), \quad &&
\text{SINR}^{zf}_{lk} =P_{lk}^{\text{zf}}\Big/\Big(I_{lk,1}^{\text{zf}}+I_{lk,2}^{\text{zf}}+I_{lk,3}^{\text{zf}}+1\Big),\\
&P_{lk}^{\text{zf}} = \rho_{dl}\left(\sum_{i\in [3]} e_{lik}\sqrt{\left(M-K_{li}\right) a_{lk}^{li} \gamma_{lk}^{li}\eta_{lik}}\right)^2, \quad &&I_{lk,1}^{\text{zf}} =  \rho_{dl}\sum_{l'\in P_l} \sum_{i\in [3]}  a^{l'i}_{lk} \left(\beta^{l'i}_{lk}-e_{l'ik}\gamma_{lk}^{l'i}\right)  \left(\sum_{k'\in \mathcal{A}_{l'i}}\eta_{l'ik'}\right),\\
&I_{lk,2}^{\text{zf}} =  \rho_{dl}\sum_{l'\not\in P_l} \sum_{i\in [3]} a^{l'i}_{lk} \beta^{l'i}_{lk}  \left(\sum_{k'\in \mathcal{A}_{l'i}}\eta_{l'ik'}\right),\quad &&I_{lk,3}^{\text{zf}} = \rho_{dl}\sum_{l'\in P_l\textbackslash\{l\}}\left(\sum_{i\in [3]} e_{l'ik}\sqrt{\left(M-K_{l'i}\right)a_{lk}^{l'i}\gamma_{lk}^{l'i}\eta_{l'ik}}\right)^2.
\end{align*}
\end{Theorem}
The proof is provided in Appendix \ref{app:thm:zf}, where the effective end to end communication channel between the antenna array $A_{li}$ and user $U_{lk}$ is modeled by a deterministic channel with additive non-Gaussian and uncorrelated noise. In Theorem~\ref{thm:zf}, $\text{SINR}^{\text{zf}}_{lk}$ can be interpreted as the \textit{effective SINR} in the sense that the capacity of an AWGN channnel with the same SINR is the same as the bound. 
Moreover, in this theorem, $P_{lk}^{\text{zf}}$ is the power of the desired signal received by $U_{lk}$, and $I_{lk,1}^{\text{zf}}$, $I_{lk,2}^{\text{zf}}$, and $I_{lk,3}^{\text{zf}}$ correspond to various types of interference experienced by the user. More specifically, $I_{lk,3}^{\text{zf}}$
is the \textit{coherent interference power} created by pilot contamination which grows linearly with the number of BS antenna elements (similar to
$P_{lk}^{\text{zf}}$). The summation of $I_{lk,1}^{\text{zf}}$ and $I_{lk,2}^{\text{zf}}$ is the \textit{non-coherent interference power} created by non-orthogonality of channel vectors of different users and channel estimation error. This type of interference does not grow with $M$, hence has negligible contribution when $M$ is large. Although increasing $M$ leads to higher SINR for all users, we remark that SINR converges to a bounded limit when $M$ goes to infinity. \textcolor{black}{We also note that the SINR expressions provided by Theorem \ref{thm:zf} does not depend on the small-scale channel coefficients which simplifies power optimization. This is the reason for performing power optimization once per large-scale coherence block.}

\subsection{Maximum-Ratio precoding} \label{subsec:mr}
In the case of using MR processing, antenna array $A_{li}$ applies precoding matrix  $\bff{B}_{li}^{\text{mr}}=\frac{1}{\sqrt{M}}\tilde{\bff{Z}}^{li^*}$ before downlink transmission. It is straightforward to show that this precoding matrix satisfies \eqref{eq:power-scale}. Furthermore, by substituting this precoding matrix in \eqref{eq:yd-1k} we have
\begin{align}
    T_1&=  
    \underbrace{\sqrt{\frac{\rho_{\text{dl}}}{M}}q_{lk}\sum_{i\in[3]}e_{lik}\sqrt{a_{lk}^{li}\gamma_{lk}^{li}\eta_{lik}}\E{\bff{z}_k^{li^T}\bff{z}_{k}^{li^*}}}_{T^{\text{mr}}_{11}}    +    \underbrace{\sqrt{\frac{\rho_{\text{dl}}}{M}}\sum_{l'\in P_l\textbackslash\{l\}}q_{l'k}\sum_{i\in[3]}e_{l'ik}\sqrt{a_{lk}^{l'i}\gamma_{lk}^{l'i}\eta_{l'ik}}\E{\bff{z}_k^{l'i^T}\bff{z}_{k}^{l'i^*}}}_{T^{\text{mr}}_{12}} \label{eq:T1-mr-expansion}\\
    &\qquad+ \underbrace{\sqrt{\frac{\rho_{\text{dl}}}{M}}\sum_{l'\in P_l}q_{l'k}\sum_{i\in[3]}e_{l'ik}\sqrt{a_{lk}^{l'i}\gamma_{lk}^{l'i}\eta_{l'ik}}\left(\bff{z}_k^{l'i^T}\bff{z}_{k}^{l'i^*}-\E{\bff{z}_k^{l'i^T}\bff{z}_{k}^{l'i^*}}\right)}_{T^{\text{mr}}_{13}}\\
    & \qquad+\underbrace{\sqrt{\frac{\rho_{\text{dl}}}{M}}\sum_{l' \in P_l}\sum_{i\in[3]}\sum_{k'\in[K]\textbackslash\{k\}}q_{l'k'}e_{l'ik}e_{l'ik'}\sqrt{a_{lk}^{l'i}\gamma_{lk}^{l'i}\eta_{l'ik'}}\bff{z}_k^{l'i^T}\bff{z}_{k'}^{l'i^*}}_{T^{\text{mr}}_{14}}
    \label{eq:yd-zf-user-sec}.
\end{align}
In this expression, $T^{\text{mr}}_{11}$ is the desired signal for user $U_{lk}$ and $T^{\text{mr}}_{12}$ represents the coherent interference caused by pilot contamination. Non-orthogonality of the user channel vectors causes interference term $T^{\text{mr}}_{14}$. Furthermore, the term $T^{\text{mr}}_{13}$ is the interference caused by \textit{beamforming gain uncertainty}. This uncertainty stems from the assumption that user $U_{lk}$ does not know the instantaneous channel state information (i.e. $\bff{z}_k^{l'i^T}\bff{z}_{k'}^{l'i^*}$) for decoding. The reason for this assumption is that it will add a large overhead to the system if the antenna arrays want to forward instantaneous channel state information to the users.
According to the term $T^{\text{mr}}_{11}$ in \eqref{eq:T1-mr-expansion}, the effective channel gain for user $U_{lk}$ is $g^{\text{eff,mr}}_{lk}=\sum_{i\in[3]} g^{\text{eff,mr}}_{lik}$
where $g^{\text{eff,mr}}_{lik}   =e_{lik}\sqrt{M\rho_{\text{dl}}a_{lk}^{li}\gamma_{lk}^{li}\eta_{lik}}$ since $\E{\bff{z}_k^{li^T}\bff{z}_{k}^{li^*}}=M$.
The following theorem provides a lower-bound on the downlink achievable rate if MR processing is applied at the antenna arrays. 

\begin{Theorem}\label{thm:mr}
For the multi-cell massive MIMO system with sectorized antennas and maximum-ratio precoding at each antenna array, the downlink transmission rate of user $U_{lk}, (l,k)\in [K]\times[L]$, $R^{\text{mr}}_{lk}$, is lower bounded by
\begin{align*}
&R^{\text{mr}}_{lk} \geq (\tau_{\text{dl}}/\tau_{\text{c}})W\log_2\left(1+\text{SINR}^{\text{mr}}_{lk}\right),\quad
&&
\text{SINR}^{mr}_{lk} =  P_{lk}^{\text{mr}}\Big/\Big(I_{lk,1}^{\text{mr}}+I_{lk,2}^{\text{mr}}+I_{lk,3}^{\text{mr}}+1\Big),\\
&P_{lk}^{\text{mr}} = \rho_{dl}\left(\sum_{i \in [3]}  e_{lik}\sqrt{Ma_{lk}^{li} \gamma_{lk}^{li} \eta_{lik}}\right)^2, \quad
&&I_{lk,1}^{\text{mr}} =  \rho_{dl} \sum_{l'\in P_l} \sum_{i\in [3]} a^{l'i}_{lk} \beta^{l'i}_{lk}  \left(\sum_{k'\in \mathcal{A}_{l'i}}\eta_{l'ik'}\right),\\
&I_{lk,2}^{\text{mr}} = \rho_{dl} \sum_{l'\not\in P_l} \sum_{i\in [3]}  a^{l'i}_{lk} \beta^{l'i}_{lk}  \left(\sum_{k'\in \mathcal{A}_{l'i}}\eta_{l'ik'}\right),\quad
&&I_{lk,3}^{\text{mr}} = \rho_{dl}\sum_{l'\in P_l\textbackslash\{l\}}\left(\sum_{i\in [3]}e_{l'ik} \sqrt{Ma_{lk}^{l'i} \gamma_{lk}^{l'i}\eta_{l'ik}}\right)^2.
\end{align*}
\end{Theorem}

The proof is provided in the Appendix \ref{app:thm:mr}. The terms introduced in this theorem has the same interpretation as the ones in Theorem \ref{thm:zf}.

\section{Power Control Strategies} \label{sec:power-opt}
Power optimization is a crucial task in massive MIMO systems by which we can improve the quality of service for the users. In this section, we investigate various power optimization methods to enhance the quality of service in a massive MIMO cellular system with sectorized antennas. 
In optimization problems of this section, we use the effective SINR expressions derived in Section \ref{sec:analysis}. As these expressions are only functions of large-scale channel coefficients and antenna directivity gains, power optimization is performed at the beginning of each large-scale coherence block. 

One of the potentials of massive MIMO systems is to provide uniformly good quality of service for the users which can be achieved through max-min throughput fairness. In this scheme, power coefficients are obtained such that the lowest effective SINR is maximized. 
In this section, we propose \textit{centralized} and \textit{decentralized} power allocation schemes to find the optimized downlink power coefficients $\eta_{lik}, (l,i,k)\in [L]\times [3] \times [K]$. 
In the centralized power allocation scheme, a central entity collects the quantities $a^{li}_{l'k}\beta^{li}_{l'k}, (l,i,l',k)\in [L]\times[3]\times[L]\times[K]$ required for power optimization to find the optimized powers. To reduce the communication overhead, we propose a decentralized approach in which the optimization problem is solved in a distributed manner.  

\subsection{Centralized Power Allocation (CPA)}\label{subsec:cpa}
\noindent \textbf{Network-wide Max-Min Fairness (NMF)}:\\
 We first consider a centralized power optimization problem under network-wide max-min fairness. This problem can be formulated as 
\begin{align}\label{max-min-sinr}
&\max_{ \{\eta_{l'ik'}\}} \; \min_{l,k} 
\left[P^{\text{s}}_{lk}\Big/\left(I^{\text{s}}_{lk,1} + I^{\text{s}}_{lk,2} + I^{\text{s}}_{lk,3} + 1\right)\right]    
\qquad\qquad\quad \\
&\text{subject to:} 
 \quad \sum_{k\in[K]} \eta_{lik} \leq 1/3, \;\;\;  (l,i)\in [L]\times[3], \notag\\
 &\qquad\qquad\quad\;\;  \eta_{lik} = 0, \;\;\;  (l,i)\in [L]\times[3],~ k \not\in \mathcal{A}_{li}, 
 \;\; \eta_{lik} \geq 0 ,\;\;\;  (l,i,k)\in [L]\times[3]\times [K], \notag
\end{align}
where, $\text{s}\in\{\text{zf},\text{mr}\}$ determines the precoding scheme. Theorems \ref{thm:zf}, \ref{thm:mr} provide $P^{\text{s}}_{lk}$, $I^{\text{s}}_{lk,1}$, $I^{\text{s}}_{lk,2}$, and $I^{\text{s}}_{lk,3}$ for ZF and MR precodings, respectively. Since the complexity of solving \eqref{max-min-sinr} depends on the communication setting, we study the problem for CoMPSec and sectorized settings separately.

\subsubsection{CoMPSec Setting with NMF}\label{subsubsec:cpa-comp}
As each user is connected to all three antenna arrays serving the corresponding cell in this setting, we have $e_{lik}=1$, $(l,i,k)\in[L]\times[3]\times[K]$. In this case, by introducing slack variables $X_{lk}$, $Y_{lk}$, and $Z_{lk}$, optimization problem (\ref{max-min-sinr}) is equivalent to:
 \begin{align}
&  \max_{ \{\psi_{l'ik'},X_{l'k'},Y_{l'k'},Z_{l'k'}\} }
\min_{l,k} 
\Big( \sum_{i\in[3]} J^{\text{s}}_{lik} \psi_{lik} \Big)^2\Big/\left( X_{lk}^2 +  Y_{lk}^2 + Z_{lk}^2+ 1\right)&& \qquad  \label{max-min-sinr-comp}\\
\text{subject to: } 
&  I^{\text{s,comp}}_{lk,1} \leq X_{lk}^2,\;\;\; (l,k)\in[L]\times[K], \notag\\
&I^{\text{s,comp}}_{lk,2} \leq Y_{lk}^2,\;\;\; (l,k)\in[L]\times[K], 
&&I^{\text{s,comp}}_{lk,3} \leq Z_{lk}^2,\;\;\; (l,k)\in[L]\times[K], \notag\\
&\sum\limits_{k\in[K]} \psi_{lik}^2 \leq 1/3  ,\;\;\;  (l,i)\in [L]\times[3], 
&&\psi_{lik} \geq 0, \;\;\; (l,i,k)\in[L]\times[3]\times[K]\notag,
\end{align}
where, $\psi_{lik}=\sqrt{\eta_{lik}}$. The explicit formulas of $J^{\text{s}}_{lik}$, $I^{\text{s,comp}}_{lk,1}$, $I^{\text{s,comp}}_{lk,2}$, and $I^{\text{s,comp}}_{lk,3}$ are presented in Table \ref{tab:comp-param}. The equivalence between (\ref{max-min-sinr}) and (\ref{max-min-sinr-comp}) follows from the fact that first three constraints in (\ref{max-min-sinr-comp}) hold with equality at the optimum.

\begin{table}[h]

\centering
\caption{Explicit formulas of the quantities used in \eqref{max-min-sinr-comp}}
\scalebox{1}{
\begin{tabular}{ccc} \toprule
$\bf Quantity$ 	             & $\bf s=zf$   & $\bf s=mr$           \\ \midrule
$J^{\text{s}}_{lik}$         & $\sqrt{\left(M-K_{li}\right)\rho_{dl} a_{lk}^{li} \gamma_{lk}^{li}}$ &  $\sqrt{M\rho_{dl} a_{lk}^{li} \gamma_{lk}^{li}}$             \\
$I^{\text{s,comp}}_{lk,1}$   & $\rho_{dl}\sum_{l'\in P_l} \sum_{i\in [3]}  a^{l'i}_{lk} \left(\beta^{l'i}_{lk}-\gamma_{lk}^{l'i}\right)  \left(\sum_{k'\in \mathcal{A}_{l'i}}\eta_{l'ik'}\right)$         &   $\rho_{dl}\sum_{l'\in P_l} \sum_{i\in [3]} a^{l'i}_{lk} \beta^{l'i}_{lk}  \left(\sum_{k'\in \mathcal{A}_{l'i}}\eta_{l'ik'}\right)$             \\
$I^{\text{s,comp}}_{lk,2}$   & $\rho_{dl}\sum_{l'\not\in P_l} \sum_{i\in [3]} a^{l'i}_{lk} \beta^{l'i}_{lk}  \left(\sum_{k'\in \mathcal{A}_{l'i}}\eta_{l'ik'}\right)$         &  $\rho_{dl}\sum_{l'\not\in P_l} \sum_{i\in [3]} a^{l'i}_{lk} \beta^{l'i}_{lk}  \left(\sum_{k'\in \mathcal{A}_{l'i}}\eta_{l'ik'}\right)$               \\
$I^{\text{s,comp}}_{lk,3}$   & $\rho_{dl}\sum_{l'\in P_l\textbackslash\{l\}}\left(\sum_{i\in [3]} \sqrt{\left(M-K_{l'i}\right)a_{lk}^{l'i}\gamma_{lk}^{l'i}}\right)^2$         &       $\rho_{dl}\sum_{l'\in P_l\textbackslash\{l\}}\left(\sum_{i\in [3]} \sqrt{Ma_{lk}^{l'i}\gamma_{lk}^{l'i}}\right)^2$         \\
\bottomrule
\end{tabular}
}
\label{tab:comp-param}

\end{table}

\begin{Lemma}
\label{lem:quasiconcavity}
Max-Min power optimization (\ref{max-min-sinr-comp}) is quasi-concave for $\text{s}\in\{\text{zf},\text{mr}\}$.
\end{Lemma}
The proof of Lemma \ref{lem:quasiconcavity} is provided in Appendix \ref{app:lem:quasiconcavity}. Due to quasi-concavity of problem (\ref{max-min-sinr-comp}), the solution can be obtained using the bisection method and a series of convex feasibility checking problems. We omit the details of this method due to the space limit and refer the reader to \cite{akbar2018downlink,shahram2018,ngo2017cellfree,shahram2017} where similar approaches are applied.





\subsubsection{Sectorized Setting with NMF}
In sectorized settings, each user is connected to one of the three antenna arrays serving the corresponding cell. As a result, only one of $e_{lik}$, $i\in[3]$ is non-zero for any $(l,k)\in [L]\times [K]$. Consequently, the terms $P_{lk}^{\text{zf}}$ and $I_{lk,3}^{\text{zf}}$ introduced in Theorem \ref{thm:zf} are simplified as follows.
\begin{align}
&P_{lk}^{\text{zf}} = \rho_{dl}\sum_{i\in [3]} \left(M-K_{li}\right) e_{lik} a_{lk}^{li} \gamma_{lk}^{li}\eta_{lik},
&I_{lk,3}^{\text{zf}} = \rho_{dl}\sum_{l'\in P_l\textbackslash\{l\}}\sum_{i\in [3]} \left(M-K_{l'i}\right) e_{l'ik}a_{lk}^{l'i}\gamma_{lk}^{l'i}\eta_{l'ik}.
\end{align}
By applying these simplifications, the numerator and denominator of $\text{SINR}^{\text{zf}}_{lk}$ provided in Theorem \ref{thm:zf}, become linear functions of the power coefficients which reduces the complexity of problem \eqref{max-min-sinr}.
Similarly, the terms $P_{lk}^{\text{mr}}$ and $I_{lk,3}^{\text{mr}}$ introduced in Theorem \ref{thm:mr} can be simplified as
\begin{align}
&P_{lk}^{\text{mr}} = \rho_{dl}\sum_{i\in [3]}M e_{lik} a_{lk}^{li} \gamma_{lk}^{li}\eta_{lik},
&I_{lk,3}^{\text{mr}} = \rho_{dl}\sum_{l'\in P_l\textbackslash\{l\}}\sum_{i\in [3]} M e_{l'ik}a_{lk}^{l'i}\gamma_{lk}^{l'i}\eta_{l'ik}.
\end{align}
By applying above simplifications, optimization problem \eqref{max-min-sinr} can be reformulated as follows.
\begin{align}
&\max_{ \{\eta_{l'ik'}\} }
\min_{l,k} \frac{ \sum_{i\in[3]} b^{\text{s}}_{lik} \eta_{lik} }{  \sum_{l'\in P_l} \sum_{i\in [3]}\sum_{k'\in \mathcal{A}_{l'i}} c^{l'i,\text{s}}_{lk}  \eta_{l'ik'} +  \sum_{l'\not\in P_l} \sum_{i\in [3]}\sum_{k'\in \mathcal{A}_{l'i}} d^{l'i,\text{s}}_{lk}  \eta_{l'ik'} + \sum_{l'\in P_l\textbackslash\{l\}}\sum_{i\in [3]} f^{l'i,\text{s}}_{lk}\eta_{l'ik}+ 1} \label{max-min-sinr-sec}\\
&\text{subject to: } \sum\limits_{k\in[K]} \eta_{lik} \leq 1/3  , (l,i)\in [L]\times[3], \notag \\ 
&\eta_{lik} = 0,  (l,i)\in [L]\times[3],~ k \not\in \mathcal{A}_{li}, \qquad \eta_{lik} \geq 0,  (l,i,k)\in[L]\times[3]\times[K]\notag.
\end{align}
Table \ref{tab:sec-param} represents the explicit formulas of $b^{\text{s}}_{lik}$, $c^{l'i,\text{s}}_{lk}$, $d^{l'i,\text{s}}_{lk}$, and $f^{l'i,\text{s}}_{lk}$ for $\text{s}\in\{\text{zf},\text{mr}\}$.

\begin{table}[h]
\centering

\caption{Explicit formulas of the quantities used in \eqref{max-min-sinr-sec}}
\scalebox{1}{
\begin{tabular}{ccc} \toprule
$\bf Quantity$ 	             & $\bf s=zf$   & $\bf s=mr$           \\ \midrule
$b^{\text{s}}_{lik}$         & $\left(M-K_{li}\right)\rho_{dl} e_{lik} a_{lk}^{li} \gamma_{lk}^{li}$         & $M\rho_{dl}e_{lik}a_{lk}^{li} \gamma_{lk}^{li}$               \\
$c^{l'i,\text{s}}_{lk}$      & $\rho_{dl} a^{l'i}_{lk} \left(\beta^{l'i}_{lk}-e_{l'ik}\gamma_{lk}^{l'i}\right) $         & $\rho_{dl} a^{l'i}_{lk} \beta^{l'i}_{lk} $  \\
$d^{l'i,\text{s}}_{lk}$      & $\rho_{dl} a^{l'i}_{lk} \beta^{l'i}_{lk}$         & $\rho_{dl} a^{l'i}_{lk} \beta^{l'i}_{lk}$ \\
$f^{l'i,\text{s}}_{lk}$      & $\left(M-K_{l'i}\right)\rho_{dl} e_{l'ik}a_{lk}^{l'i}\gamma_{lk}^{l'i}$         & $M\rho_{dl}e_{l'ik}a_{lk}^{l'i} \gamma_{lk}^{l'i}$               \\
\bottomrule
\end{tabular}
}
\label{tab:sec-param}

\end{table}

\begin{Lemma}
\label{lem:quasilinearity}
Power optimization problem (\ref{max-min-sinr-sec}) is quasi-linear for $\text{s}\in\{\text{zf},\text{mr}\}$.
\end{Lemma}

The proof is followed by the linearity of constraints as well as the linearity of numerator and denominator of the objective function (SINR) with respect to the downlink power coefficients. As a result of Lemma \ref{lem:quasilinearity}, a bisection search can lead to the optimal solution of problem \eqref{max-min-sinr-sec}. 

It is well-known that when max-min fairness is considered, the network performance is restricted by the users experiencing the lowest SINR due to severe shadowing and/or high interference. The reason is that in order to enhance the effective SINR of such users, the power optimization algorithm decreases the transmit power of other users to reach a balanced SINR for all users. Therefore, considering network-wide max-min fairness can lead to a low network performance (e.g. low sum-rate) as the number of cells $L$ increases. To address this issue, we consider per-cell max-min fairness where the objective is to equalize the effective SINRs within each cell. An additional upside of using per-cell max-min fairness is that it allows us to derive closed-form expressions for downlink power coefficients. Hence, power optimization can be performed with very low computational complexity.  \\

\noindent \textbf{Per-cell Max-Min Fairness (PMF)}:\\
In this case, the centralized entity solves the power optimization problem separately for each cell, assuming that the coherent interference is negligible and all the antenna arrays transmit at maximum power. Specifically, we assume that: (I) \textit{each antenna array transmits with maximum power}, i.e. $\sum_{k\in[K]} \eta_{lik}=1/3$, $(l,i)\in [L]\times [3]$, and (II) \textit{non-coherent interference is negligible}, i.e. $I^{\text{s}}_{lk,3} \approx 0$,  $\text{s}\in\{\text{zf},\text{mr}\}$.
Using these assumptions, it can be shown that the effective SINR of the users belonging to the cell $C_l$ do not depend on the power coefficients used in the cells other than $C_l$. This in turn enables performing power optimization on a cell basis. Assumption (I) can be satisfied as the power coefficients are under control. Furthermore, the simulation results in Section \ref{sec:sim} confirm that assumption (II) is valid in practical scenarios with moderately high number of antennas. Next, we study the problem for CoMPSec and sectorized settings separately.

\subsubsection{CoMPSec Setting with PMF}
\label{subsubsec:comp-pmf}
Considering SINR expressions of CoMP setting together with assumptions (I) and (II), we can formulate the power optimization for cell $C_l$, $l\in [L]$ as follows.
\begin{align}
&\max_{ \{\eta_{lik'}\} }
\min_{k\in[K]} 
\Big[ \Big(\sum_{i \in [3]} m^{\text{s}}_{lik} \sqrt{\eta_{lik}}\Big)^2\Big/  \Big(\sum_{l'\in P_l} \sum_{i\in [3]} n^{l'i,\text{s}}_{lk}/3 +  \sum_{l'\not\in P_l} \sum_{i\in [3]} o^{l'i,\text{s}}_{lk}/3   + 1\Big)\Big]\label{max-min-sinr-comp-local}\\
&\text{subject to: } \sum\limits_{k\in[K]} \eta_{lik} = 1/3  ,\;  i\in [3],\;\;\eta_{lik} \geq 0, \;(i,k)\in [3]\times[K]\notag.
\end{align}
where $m^{\text{s}}_{lik}$, $n^{l'i,\text{s}}_{lk}$, and $o^{l'i,\text{s}}_{lk}$ are listed is Table \ref{tab:comp-param-pmf}. Note that the centralized entity has to solve \eqref{max-min-sinr-comp-local} for each $l\in [L]$ separately. However, the problem can be solved for all the cells in parallel. It can be shown that \eqref{max-min-sinr-comp-local} is equivalent to a quasi-concave problem which can be solved by using bisection algorithm. However, we can derive a closed form solution for problem \eqref{max-min-sinr-comp-local} by adding another constraint to the problem. More specifically, if we assume that the power coefficients assigned to user $U_{lk}$ by the antenna arrays $A_{li}$ are equal, i.e. $\eta_{l1k}=\eta_{l2k}=\eta_{l3k}\triangleq \eta_{lk}$, then we can solve problem \eqref{max-min-sinr-comp-local} analytically. Under this assumption, we can reformulate problem \eqref{max-min-sinr-comp-local} as 
\begin{align}
&\left\{\eta^{*,\text{s}}_{lk'}\right\}_{k'\in [K]}=\argmax_{ \{\eta_{lk'}\} }
\min_{k\in[K]} \frac{ r^{\text{s}}_{lk}\eta_{lk} }{  \frac{1}{3}\sum_{l'\in P_l} \sum_{i\in [3]} n^{l'i,\text{s}}_{lk} +  \frac{1}{3}\sum_{l'\not\in P_l} \sum_{i\in [3]} o^{l'i,\text{s}}_{lk}   + 1} \label{max-min-sinr-comp-local2}\\
&\text{subject to: } \sum\limits_{k\in[K]} \eta_{lk} = 1/3  ,\;\;\;\eta_{lk} \geq 0, \;  k\in [K]\notag.
\end{align}
where $\text{s}\in\{\text{zf},\text{mr}\}$ and $r^{\text{s}}_{lk}$ is presented in Table \ref{tab:comp-param-pmf}. Lemma \ref{lem:comp-pmf} provides the solution to \eqref{max-min-sinr-comp-local2}.

\begin{Lemma}\label{lem:comp-pmf}
The solution to problem \eqref{max-min-sinr-comp-local2} is 
\begin{align}
\eta^{*,\text{s}}_{lk}=\left(D^{\text{s}}_{lk}/r^{\text{s}}_{lk}\right)~\overline{\text{SINR}}^{\text{s}}_l,\;\; (k,l) \in [K]\times [L],
\end{align}
where $D^{\text{s}}_{lk}=\frac{1}{3}\sum_{l'\in P_l} \sum_{i\in [3]} n^{l'i,\text{s}}_{lk} +  \frac{1}{3}\sum_{l'\not\in P_l} \sum_{i\in [3]} o^{l'i,\text{s}}_{lk} + 1$ and $\overline{\text{SINR}}^{\text{s}}_l=\left(3\sum_{k'\in[K]}(D^{\text{s}}_{lk'}/r^{\text{s}}_{lk'})\right)^{-1}$.
\end{Lemma}
The proof is provided in Appendix \ref{app:lem:sec-pmf}, where we show that using $\eta^{*,\text{s}}_{lk}$ leads to the same effective SINR for all the users in each cell since otherwise the minimum SINR can be increased in that cell. $\overline{\text{SINR}}^{\text{s}}_l$ in Lemma \ref{lem:comp-pmf} represents the common effective SINR in cell $C_l$ when using the optimized power coefficients for precoding scheme $\text{s}\in\{\text{zf},\text{mr}\}$.

\begin{table}[h]
\centering

\caption{Explicit formulas of the quantities used in \eqref{max-min-sinr-comp-local} and \eqref{max-min-sinr-comp-local2}}
\scalebox{1}{
\begin{tabular}{ccc} \toprule
$\bf Quantity$ 	             & $\bf s=zf$   & $\bf s=mr$           \\ \midrule
$m^{\text{s}}_{lik}$         & $\sqrt{(M-K_{li})\rho_{dl}a_{lk}^{li} \gamma_{lk}^{li}}$         & $\sqrt{M\rho_{dl}a_{lk}^{li} \gamma_{lk}^{li}} $               \\
$n^{l'i,\text{s}}_{lk}$      & $\rho_{dl} a^{l'i}_{lk} \left(\beta^{l'i}_{lk}-\gamma_{lk}^{l'i}\right) $         & $\rho_{dl} a^{l'i}_{lk} \beta^{l'i}_{lk} $  \\
$o^{l'i,\text{s}}_{lk}$      & $\rho_{dl} a^{l'i}_{lk} \beta^{l'i}_{lk}$         & $\rho_{dl} a^{l'i}_{lk} \beta^{l'i}_{lk}$ \\
$r^{\text{s}}_{lk}$      & $\left(\sum_{i \in [3]} \sqrt{(M-K_{li})\rho_{dl}a_{lk}^{li} \gamma_{lk}^{li}} \right)^2$         & $\left(\sum_{i \in [3]} \sqrt{M\rho_{dl}a_{lk}^{li} \gamma_{lk}^{li}} \right)^2$               \\
\bottomrule
\end{tabular}
}
\label{tab:comp-param-pmf}

\end{table}

\subsubsection{Sectorized Setting with PMF}
Using the SINR expressions of sectorized setting together with assumptions (I) and (II), power optimization for cell $C_l$ can be formulated as follows.
\begin{align}
&\left\{\eta^{*,\text{s}}_{lik'}\right\}_{(i,k')\in [3]\times[K]}=\argmax_{ \{\eta_{lik'}\} }
\min_{k\in[K]} \frac{ \sum_{i\in[3]} b^{\text{s}}_{lik} \eta_{lik} }{  \frac{1}{3}\sum_{l'\in P_l} \sum_{i\in [3]} c^{l'i,\text{s}}_{lk}   +  \frac{1}{3}\sum_{l'\not\in P_l} \sum_{i\in [3]}d^{l'i,\text{s}}_{lk}  + 1} \label{max-min-sinr-sec-local}\\
&\text{subject to: } \sum\limits_{k\in[K]} \eta_{lik} \leq 1/3, \; i\in [3]  ,\;\; \eta_{lik} = 0, \; k \not\in \mathcal{A}_{li},  \; i\in [3],\;\;\;\eta_{lik} \geq 0, \; (i,k)\in [3]\times[K]\notag.
\end{align}
 where $b^{\text{s}}_{lik}$, $c^{l'i,\text{s}}_{lk}$, and $d^{l'i,\text{s}}_{lk}$ are listed in Table \ref{tab:sec-param}. Following lemma provides closed-form solution to this optimization problem. 

\begin{Lemma}\label{lem:sec-pmf}
The solution to problem \eqref{max-min-sinr-sec-local} is 
\begin{align}
\eta^{*,\text{s}}_{lik}=
\begin{cases}
\frac{E^{\text{s}}_{lk}}{F^{\text{s}}_{lk}}~\overline{\text{SINR}}^{\text{s}}_{li}~~~ \text{if  $k \in \mathcal{A}_{li}$},\\
0 ~~~~~~~~~~~~~ \text{if  $k \not\in \mathcal{A}_{li}$},
\end{cases}
(k,i,l) \in [K]\times [3]\times [L]
\end{align}
where $E^{\text{s}}_{lk}=\frac{1}{3}\sum_{l'\in P_l} \sum_{i\in [3]} c^{l'i,\text{s}}_{lk} +  \frac{1}{3}\sum_{l'\not\in P_l} \sum_{i\in [3]} d^{l'i,\text{s}}_{lk} + 1$, $F^{\text{s}}_{lk}=\sum_{i\in[3]} b^{\text{s}}_{lik}$, and $\overline{\text{SINR}}^{\text{s}}_{li}=\left(3\sum_{k'\in\mathcal{A}_{li}}\frac{E^{\text{s}}_{lk'}}{F^{\text{s}}_{lk'}}\right)^{-1}$.
\end{Lemma}
The proof is omitted due the space limit. A sketch of the proof is as follows. We first show that problem \eqref{max-min-sinr-sec-local} can be decomposed into three independent optimization problems one for each antenna array serving cell $C_l$. Then, we use a similar approach as the one provided in the proof of Lemma \ref{lem:comp-pmf} to obtain the closed-form solution to these three problems.


\subsection{Decentralized Power Allocation (DPA)} \label{subsec:dpa}
\textcolor{black}{The CPA strategies described in Section \ref{subsec:cpa} require a central entity collecting the quantities $a^{li}_{l'k}\beta^{li}_{l'k}, (l,i,l',k)\in [L]\times[3]\times[L]\times[K]$ from the entire network at each large-scale coherence block. However, this may lead to a high communication overhead. In this section, we propose a decentralized power allocation scheme where the power coefficients are optimized in a distributed manner. To elaborate, let $\mathcal{V}_l$ denote the set of cells which are direct neighbors to cell $C_l$. In DPA, antenna array $A_{li}$ considers itself and all the antenna arrays in $\mathcal{V}_l$ such that if $\mathcal{V}_l$ would be the entire network, overlooking the existence of the rest of the cells. $A_{li}$ collects all the quantities $a^{li}_{l'k}\beta^{li}_{l'k}$ in $\mathcal{V}_l$ and solves the optimization problem formulated for this network using the same approaches described for CPA in this section. Consequently, $A_{li}$ can find all the power coefficients in $\mathcal{V}_l$. Next, $A_{li}$ uses the found downlink power coefficients $\eta_{lik}, k\in [K]$, and discard the powers found for other antenna arrays in $\mathcal{V}_l$. In DPA, the optimization problem solved by each antenna array can be based on NMF or PMF criterion. Similar approaches as the ones used for CPA (described in Section \ref{subsec:cpa}) can be used to solve these problems. Table \ref{tab:comparison} provides a comparison among all the proposed power allocation schemes in terms of computational complexity and communication overhead. As shown in this table, using DPA along with PMF is the most appealing choice as its overhead and complexity is low.}   
\begin{table}[h]
\centering
\caption{A comparison among the proposed power allocation schemes} 
\scalebox{1}{
\begin{tabular}{cccc} \toprule
\textbf{Method} & \textbf{Fairness} & \textbf{Overhead} & \textbf{Complexity}\\\toprule
 Centralized		 & \begin{tabular}{l} Network-wide max-min \\ Per-cell max-min \end{tabular} & \begin{tabular}{l} High \\ High \end{tabular} & \begin{tabular}{l} High \\ Low \end{tabular} \\ \hline
Decentralized		 & \begin{tabular}{l} Network-wide max-min \\ Per-cell max-min \end{tabular} & \begin{tabular}{l} Low \\ Low \end{tabular} & \begin{tabular}{l} High \\ Low \end{tabular} \\
\bottomrule
\end{tabular}
}
\label{tab:comparison}

\end{table}
\section{Simulation Results} \label{sec:sim}
In this section, we provide extensive simulation results to evaluate the performance of the proposed settings. We consider a network composed of $L = 19$ cells (two rings of cells around a central cell), each with a radius of $R=1$ km. We assume that $K = 18$ users are distributed uniformly within each cell except for a disk with radius $r = 50$ m around the BSs. The cells are wrapped to avoid the cell edge effect.
Moreover, we consider pilot reuse factors $\xi=1,3$ in the simulations. Table \ref{tab:sim-param} lists the simulation parameters. As a benchmark, we consider an omnidirectional setting (referred to as `Omni' setting) in which a BS consisting of $M_{BS}=3M$ omnidirectional antenna elements is placed at the center of each cell and serves every user in that cell. Furthermore, to evaluate the impact of the power control schemes described in Section \ref{sec:power-opt}, we consider a uniform power allocation (UPA) scheme in which each array transmits with full power and divides this power uniformly among the users connected to that array.

\begin{table}[h]
\centering

\caption{Simulation parameters}
\scalebox{1}{
\begin{tabular}{llll} \toprule
$\bf Parameter$ 				 & $\bf Value$  &$\bf Parameter$ 	 & $\bf Value$             \\ \midrule
Path loss model        		     & COST231                 
&Frequency                        & $1900$ MHz              \\
Bandwidth $W$                    & $20$ MHz                
&Cell radius                      & $1$ Km                 \\
Number of cells $L$              & $19$               
&Number of users per cell $K$     & $18$ (from \cite{marzetta2016fundamentals})               \\
Number of antennas per array $M$ & $100$                
&Pilot reuse factor $\xi$              & $1,3$                \\
Thermal noise spectral density   & $-174$ dBm/Hz                
&Noise figure                     & $9$ dB \\
Maximum BS transmit power        & $30$ dBm                
&Maximum user transmit power      & $23$ dBm                \\
Coherence bandwidth $B_c$        & $210$ KHz           
&Coherence time $T_c$             & $2$ msec           \\
Number of pilots $\tau_p$        & $K\xi$           
&Downlink to uplink resource share $\tau_{\text{dl}}/\tau_{\text{ul}}$  & $2$ (from \cite{marzetta2016fundamentals})   \\ 
\bottomrule
\end{tabular}
}
\label{tab:sim-param}

\end{table}

\subsection{Impact of sectorization} \label{subsec:sec-effect}
\begin{figure}[h]
    \centering
        \begin{subfigure}[b]{0.49\textwidth}
            \includegraphics[width=\textwidth]{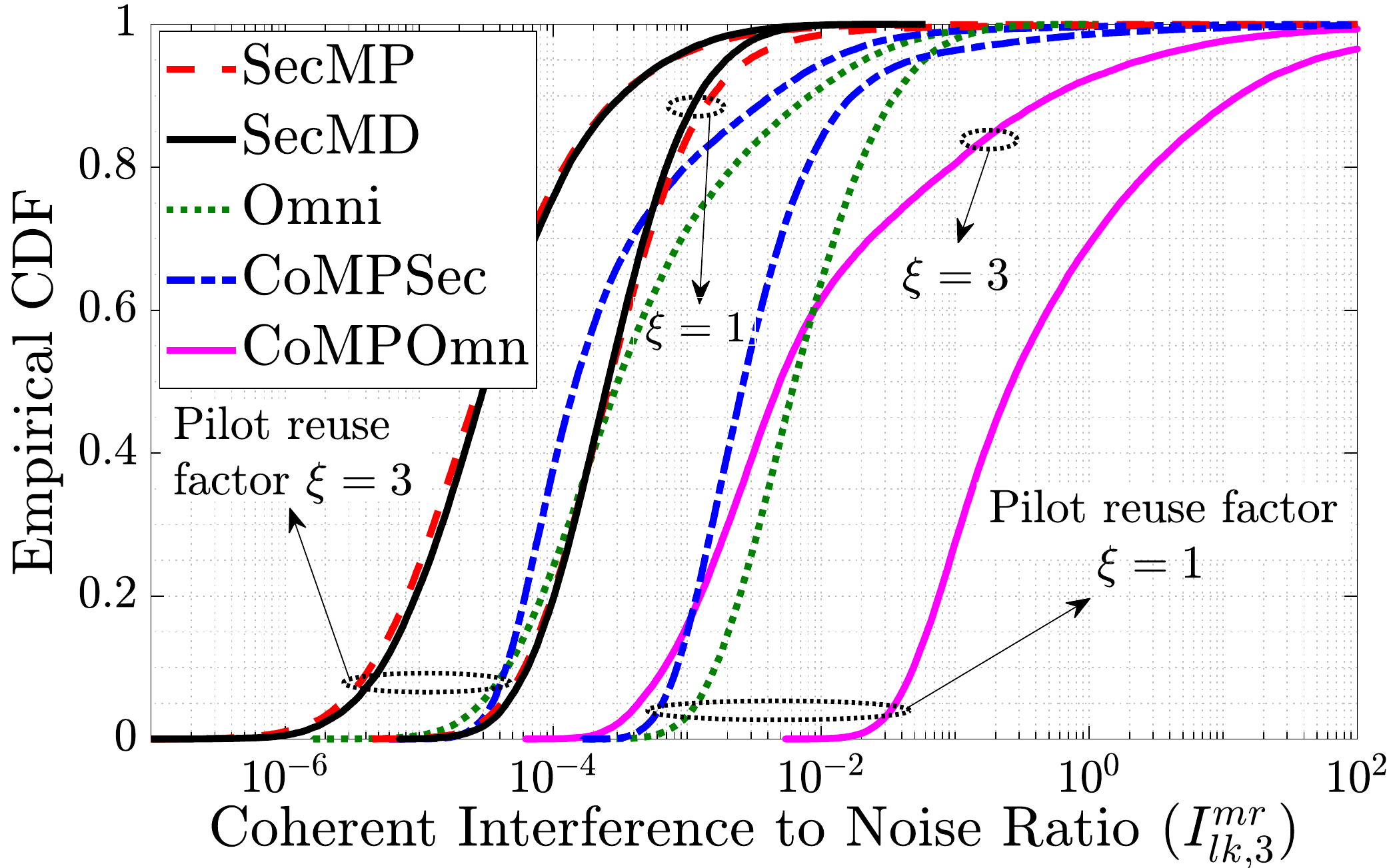}
            \caption[]%
            {{\small }}    
            \label{fig:coh-inr}
        \end{subfigure}
        \begin{subfigure}[b]{0.49\textwidth}              \includegraphics[width=\textwidth]{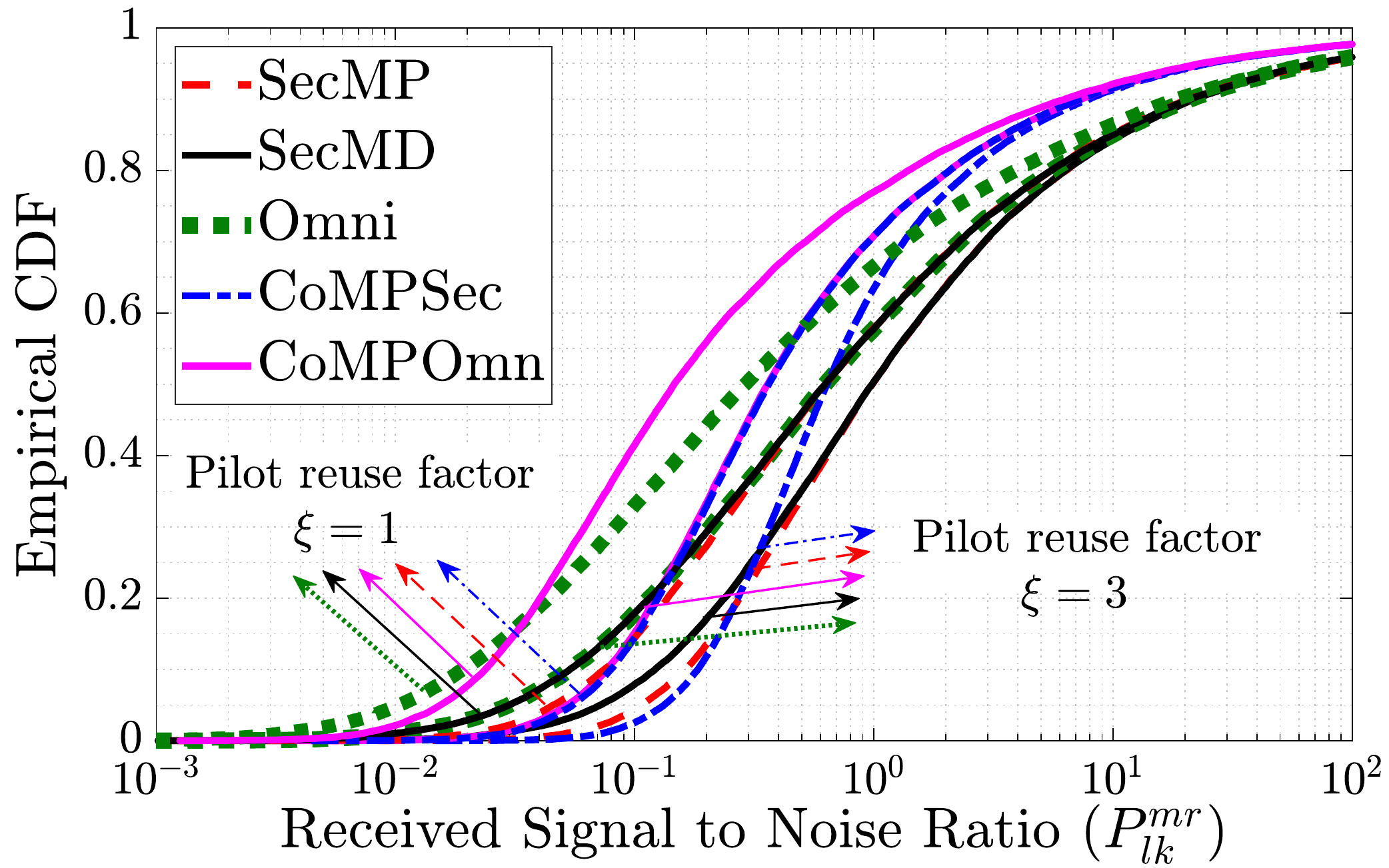}
            \caption[]%
            {{\small}}    
            \label{fig:snr}
        \end{subfigure}
        
        \caption{(a) Empirical CDF of coherent interference power to noise ratio, (b) Empirical CDF of desired signal power to noise ratio.}
        \label{fig:sec-pilot-effect}

\end{figure}

\begin{figure}[h]
    \centering
    \includegraphics[width=0.49\textwidth]{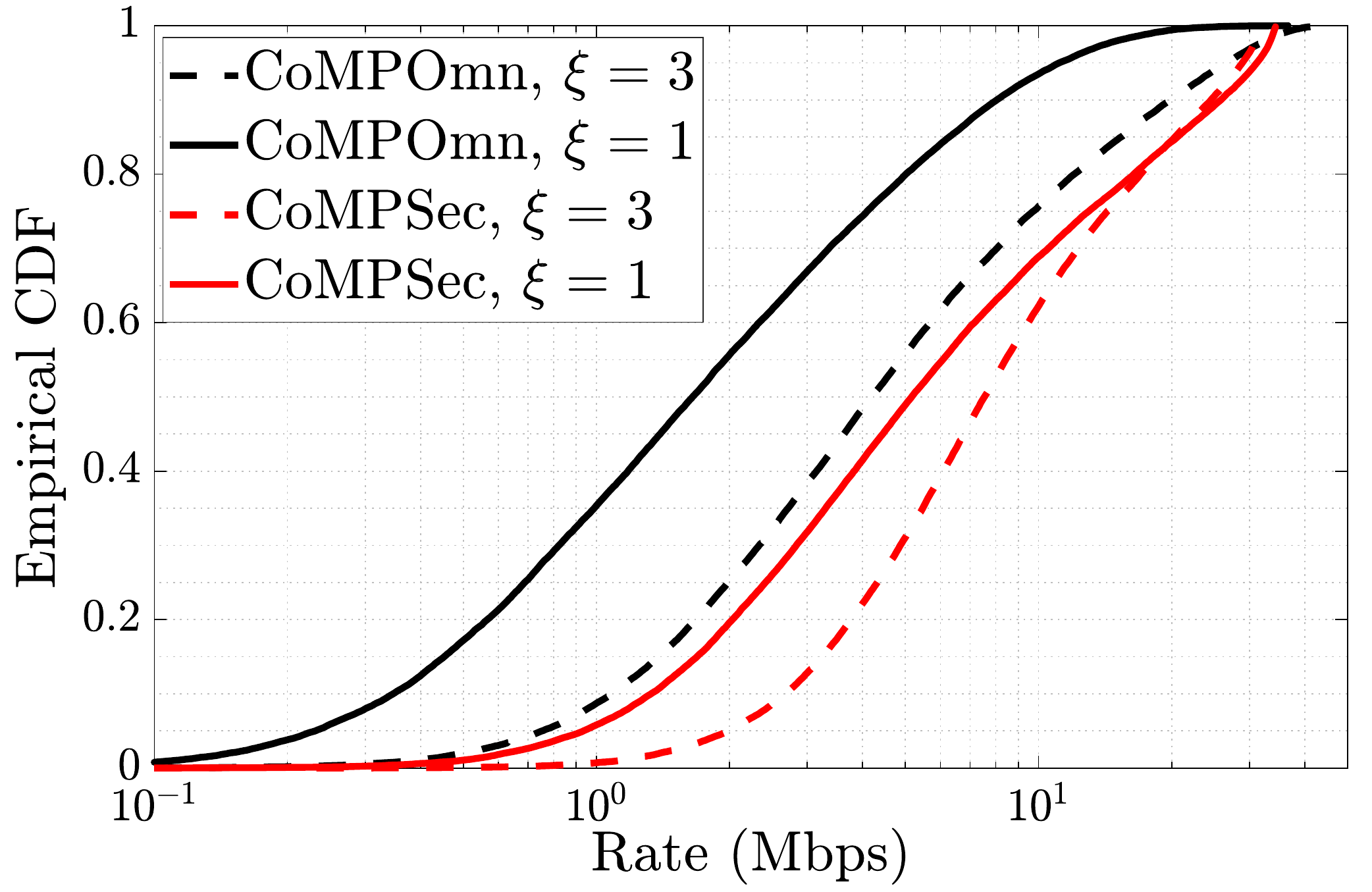}
    \caption{Empirical CDF of users' achievable rate for CoMPSec and CoMPOmn settings with $\xi=1,3$ when MR precoding is applied.}
    \label{fig:CDF-comp-rate}
\end{figure}

In this section, we investigate the effect of using sectorized antennas in the system. To evaluate the ultimate gains of sectorization, we adopt IRP model described in Section \ref{subsec:antenna} with parameters $\theta=120^\circ$, $a_{\text{Q}}=3$, and $a_{\text{q}} = 0$. Furthermore, as power optimization affects the performance, we consider UPA in this section to identify the impact of sectorization apart from power optimization.  Figure \ref{fig:sec-pilot-effect} illustrates the coherent interference power as well as the desired received signal power both normalized to the noise power for different communication settings when MR precoding is applied. Comparing omnidirectional setting (Omni) with SecMD and SecMP settings in Figure \ref{fig:sec-pilot-effect}, there are two important observations for both pilot reuse factors: \textit{i)} the coherence interference power in sectorized settings (i.e. SecMD and SecMP) is significantly lower than that in the Omni setting and \textit{ii)} the desired received signal power is higher in sectorized settings than Omni setting. The main reason behind theses observations is that the quality of channel estimation is higher when employing sectorized antenna elements. In the omnidirectional setting, each antenna receives the pilots transmitted from all cells and in all directions. However, sectorized antennas receive these signals with different directionality gains from the users in different cells. More precisely, one-third of the signals (those in the main-lobe coverage of the antenna) including the desired pilot are amplified with $a_{\text{Q}} = 3$, while the remaining two-thirds (in the back-lobe coverage of the antenna) are attenuated with $a_{\text{q}} = 0$. 
In this case, the effective channel estimation SINR is approximately three times larger compared to the Omni setting, which in turn reduces the impact of pilot contamination. We observed similar trends when ZF precoding is applied. Another key observation form Figure \ref{fig:coh-inr} is that coherent interference power is negligible except for CoMPOmn setting which will be discussed later. This implies that when the number of antenna elements is moderately high, the main effect of pilot contamination is to reduce the desired received power. Furthermore, we note that increasing pilot reuse factor from one to three reduces pilot contamination effect since the number of interfering pilots is one-third with $\xi=3$. \textcolor{black}{Comparing CoMPSec and CoMPOmn settings in Figure \ref{fig:sec-pilot-effect}, we observe that using omnidirectional antennas intensifies the impact of pilot contamination significantly when pilots are reused at every cell, i.e. when $\xi=1$. The reason is that for a given pilot sequence in this case (i.e., CoMPOmn with $\xi=1$), a BS located at the corner of three neighboring cells should serve three users, one in each cell, all using the same pilot which causes severe pilot contamination and channel estimation loss. One approach to alleviate pilot contamination in this case is to not reuse the pilots in every cell by taking $\xi=3$. However, this leads to an extra overhead for the system as longer pilot sequences are required to serve the same number of users. A better solution is to use sectorized antennas instead of omnidirectional ones (as in CoMPSec setting) which reduces pilot contamination by filtering the interfering pilots in the spatial domain without increasing the overhead. Figure \ref{fig:CDF-comp-rate} depicts the empirical CDF of users' achievable rate in CoMPOmn and CoMPSec for $\xi=1,3$. We observe that using CoMPSec even with $\xi=1$ outperforms CoMPOmn with both $\xi=1,3$. Therefore, between the two settings with multi-point coordination, we will only consider CoMPSec in the rest of this section due to its superiority.}


\subsection{Achievable rate}
\begin{figure}[h]
    \centering
        \begin{subfigure}[b]{0.49\textwidth}
            \includegraphics[width=\textwidth]{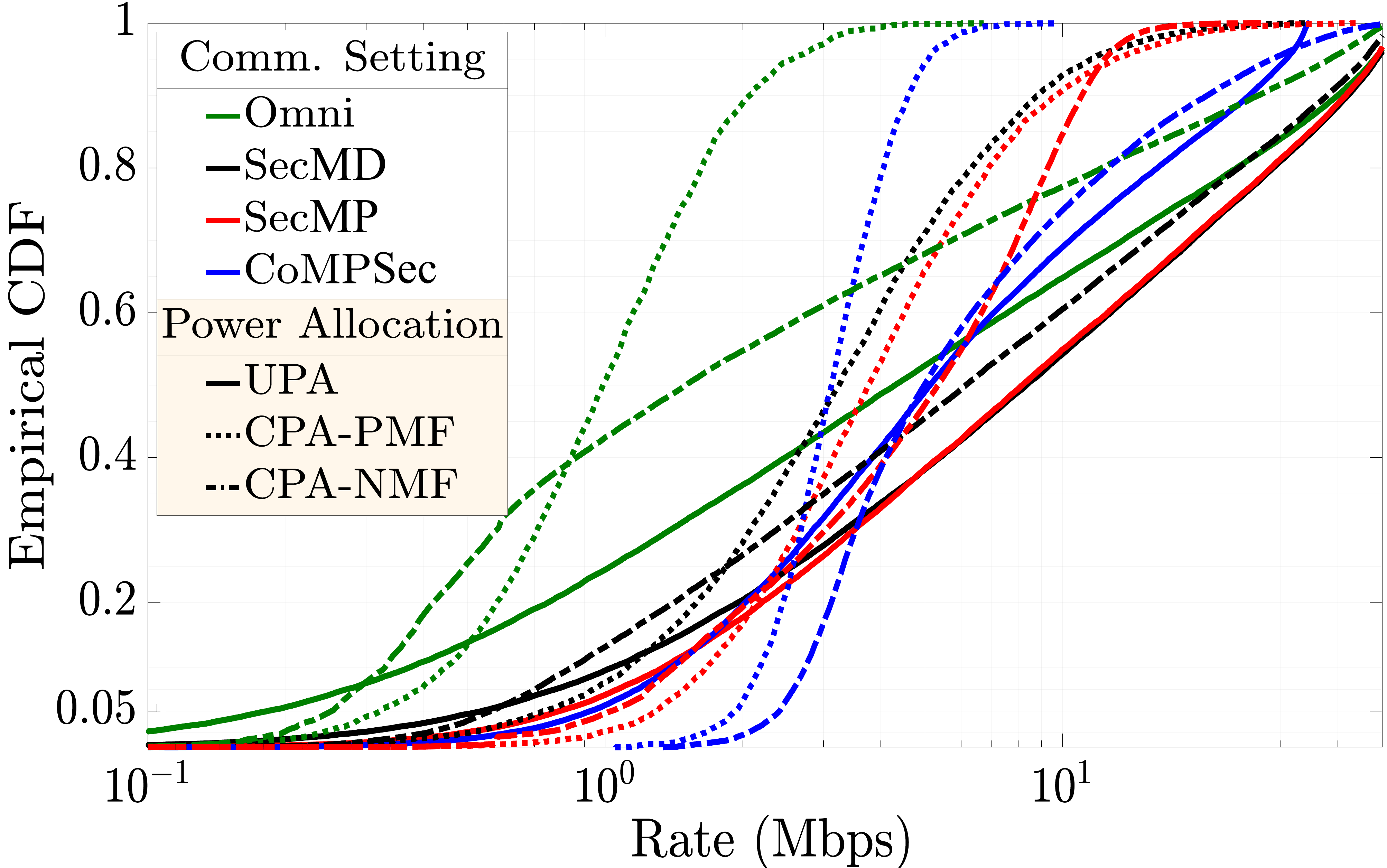}
            \caption{Maximum-ratio precoding}%
            {{\small }}    
            \label{fig:rateCDF-mr}
        \end{subfigure}
        \begin{subfigure}[b]{0.49\textwidth}              \includegraphics[width=\textwidth]{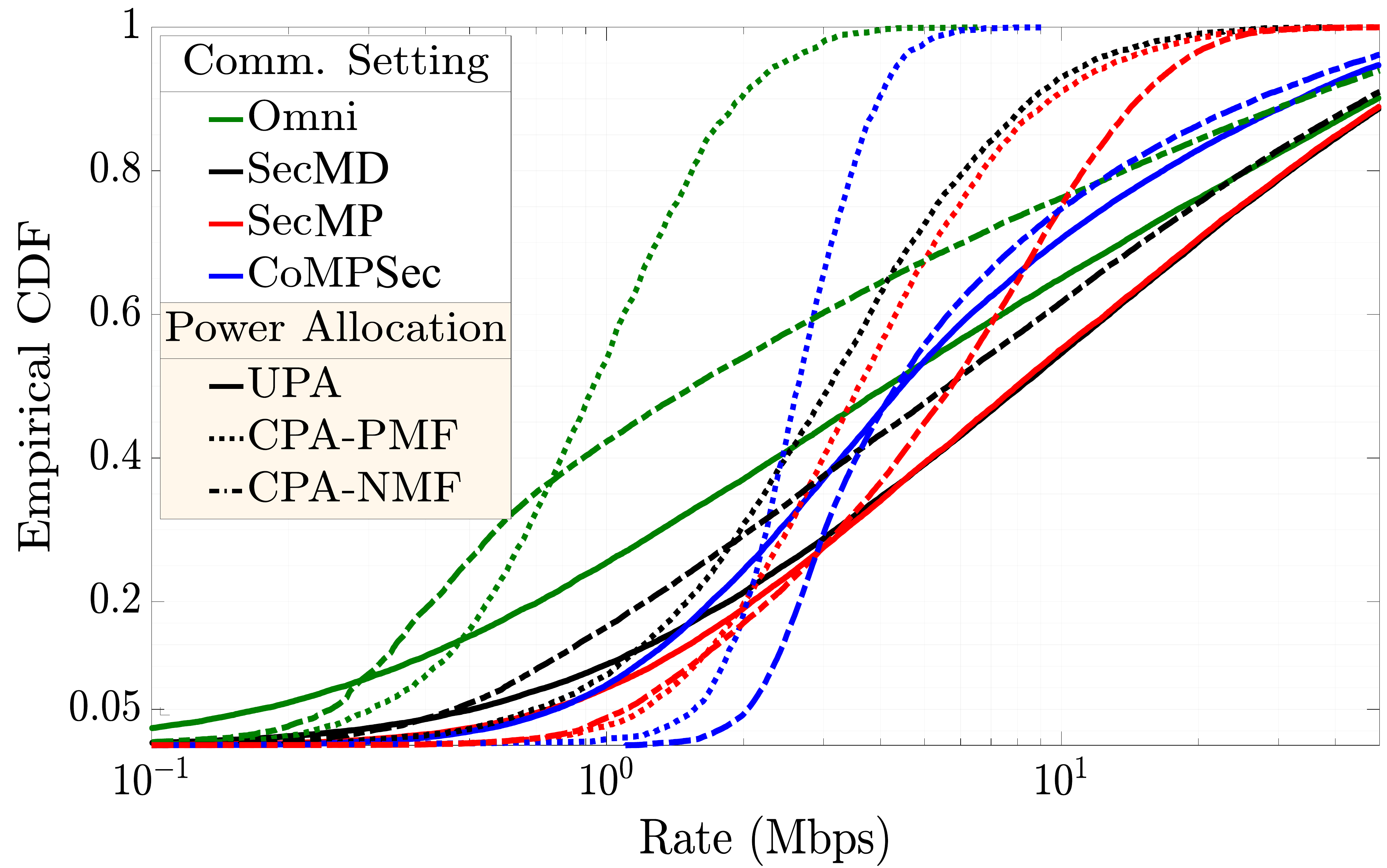}
            \caption{Zero-forcing precoding}%
            {{\small}}    
            \label{fig:rateCDF-zf}
        \end{subfigure}
        
        \caption{Empirical CDF of downlink achievable rate with pilot reuse factor $\xi=1$ for different precoding, communication setting, and power allocation schemes.}

         \label{fig:rateCDF}
\end{figure}

In this section, we investigate the impact of sectorization and multi-point coordination on the downlink achievable rate in massive MIMO systems. To compare different settings, we use median rate as well as $95\%-likely$ rate defined as the rate which is achievable by $95\%$ of the users in the network \cite{marzetta2016fundamentals}. We consider IRP model with parameters $\theta=120^\circ$, $a_{\text{Q}}=3$, and $a_{\text{q}} = 0$ to assess the ultimate impact of sectorization on the rates. Figure \ref{fig:rateCDF} shows the empirical CDF of the downlink achievable rate (i.e. lower-bound introduced in Theorems \ref{thm:zf} and \ref{thm:mr}) for different communication settings and various centralized power allocation schemes when pilot reuse factor is one. Considering $95\%-likely$ rate, we have the following observations:
\begin{enumerate}[label=\textit{\roman*}),wide = 0pt]
    \item  Sectorized settings (SecMD, SecMP) outperform omnidirectional setting due to lower pilot contamination effect discussed in Section \ref{subsec:sec-effect}. 
    \item SecMP outperforms SecMD setting. The reason is that in SecMP, the user is connected to the array from which it receives the highest average power wherease in SecMD, it connects to the closest array from which it may receive a lower power due to the effect of shadowing. This leads to a type of \textit{selection diversity} in SecMP which enhances the quality of service as expected.
    \item \textcolor{black}{CoMPSec further increases the $95\%-likely$ rate by leveraging \textit{macro-diversity}. The reason is that the resulting power gain improves the quality of service for users with weaker communication channels more compared to others as they are in a power limited regime.}
    \item Max-min power control improves the quality of service for the low-SINR users in the system as expected. We observe that per-cell max-min power control leads to acceptable performance while reducing the computational complexity significantly.
    \textcolor{black}{\item While $95\%-likely$ rate is higher with PMF compared to NMF in most of the settings, NMF outperforms PMF in terms of $95\%-likely$ rate in the CoMPSec setting. The major reason for this is the extra assumption made to derive the closed form solution to the PMF for CoMPSec in Section \ref{subsubsec:comp-pmf}. This assumption restricts the power coefficients $\eta_{lik}, i\in[3]$ to be equal which reduces the complexity by allowing us to obtain a closed form solution, at the cost of a slight performance loss.}
\end{enumerate} 

Table \ref{tab:comparison-ideal} lists the $95\%-likely$ rate as well as the median rate for different communication settings when MR precoding and CPA-PMF are applied. Considering omnidirectional setting as the benchmark, we observe that sectorization and multi-point coordination lead to significant improvements in the $95\%-likely$ rate. Furthermore, sectorized settings (SecMP and SecMD) provide higher median rates than CoMPSec setting when max-min fairness is applied. \textcolor{black}{This observation is not surprising as the power optimization objective is to maximize the per-cell minimum rate and the extra degrees of freedom in CoMPsec have been used to improve the minimum rate. Consequently, CoMPSec provides a higher $95\%-likely$ rate while leading to a lower median rate compared to SecMP and SecMD.}

\begin{table}[t]
\centering
\caption{ $95\%-likely$ rate and median rate when MR precoding, CPA-PMF, and ideal radiation pattern are considered. The number in parentheses is the gain compared to the benchmark setting.}
\scalebox{1}{
\begin{tabular}{ccll} \toprule
\begin{tabular}{c}\textbf{Pilot reuse}\\\textbf{factor}\end{tabular}&\begin{tabular}{c}\textbf{Communication}\\\textbf{Setting}\end{tabular} & \begin{tabular}{c}$\bf 95\%-likely$\\$\bf rate$ (Mbps)\end{tabular} & \begin{tabular}{l} $\bf Median~rate$\\~~~~(Mbps) \end{tabular} \\\toprule
$\xi=1$ & \begin{tabular}{l} \textbf{Omni}\\ \textbf{SecMD}\\\textbf{SecMP}\\\textbf{CoMP} \end{tabular} & \begin{tabular}{l} 0.31 (benchmark)\\ 0.73 ($2.35\times$)\\1.29 ($4.16\times$)\\1.94 ($6.25\times$) \end{tabular} &\begin{tabular}{l} 0.98 (benchmark)\\ 3.24 ($3.30\times$)\\3.76 ($3.83\times$)\\3.12 ($3.18\times$) \end{tabular}\\ 
\bottomrule
$\xi=3$ & \begin{tabular}{c} \textbf{Omni}\\ \textbf{SecMD}\\\textbf{SecMP}\\\textbf{CoMP} \end{tabular}		 & \begin{tabular}{l} 0.84 (benchmark)\\ 1.61 ($1.91\times$)\\2.63 ($3.13\times$)\\3.79 ($4.51\times$) \end{tabular} & \begin{tabular}{l} 2.38 (benchmark)\\ 5.57 ($2.34\times$)\\6.26 ($2.63\times$)\\5.49 ($2.30\times$) \end{tabular}   \\ 
\bottomrule
\end{tabular}
}
\label{tab:comparison-ideal}

\end{table}

\begin{figure}[h]
    \centering
        \begin{subfigure}[b]{0.5\textwidth}
            \includegraphics[width=\textwidth]{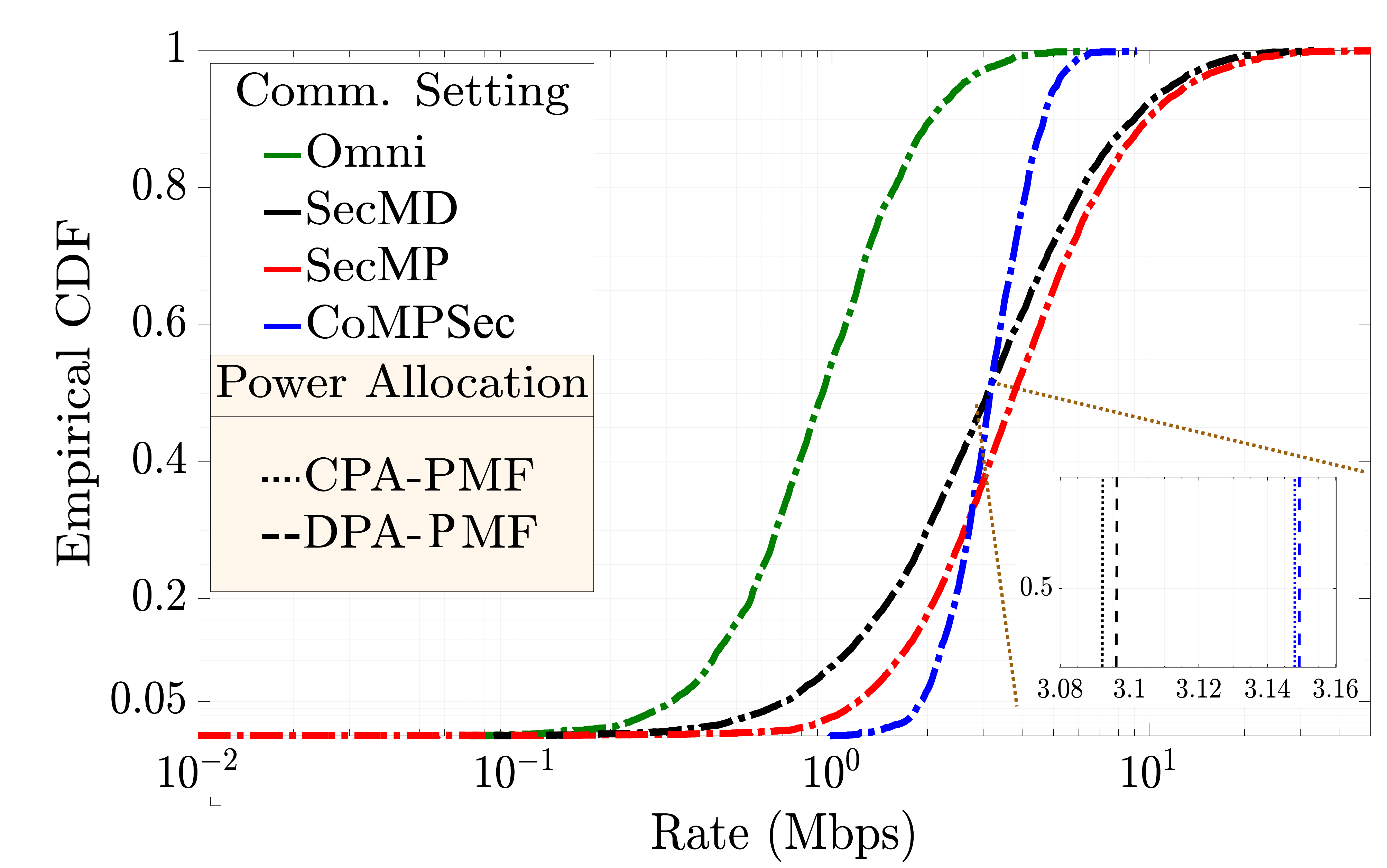}
            \caption{Maximum-ratio precoding}%
            {{\small }}    
            \label{fig:rateCDF-power-mr}
        \end{subfigure}
        \begin{subfigure}[b]{0.49\textwidth}              \includegraphics[width=\textwidth]{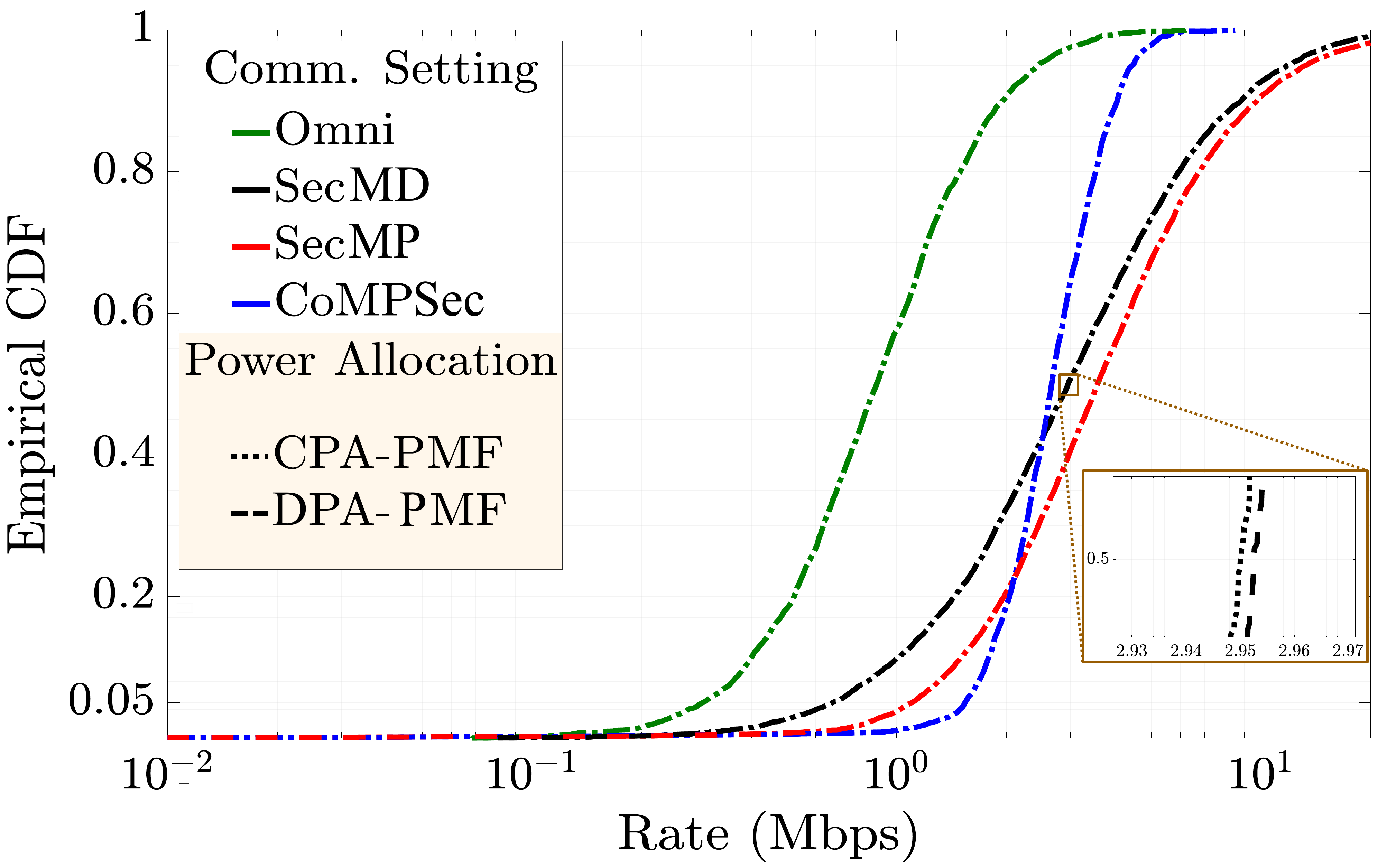}
            \caption{Zero-forcing precoding}%
            {{\small}}    
            \label{fig:rateCDF-power-zf}
        \end{subfigure}
        
        \caption{Comparison of centralized and decentralized power allocation strategies with per-cell max-min fairness criterion and pilot reuse factor $\xi=1$.}

         \label{fig:rateCDF-power}
          
\end{figure}

We investigate the performance of the decentralized power allocation in Figure \ref{fig:rateCDF-power}. In this figure, we consider centralized and decentralized schemes with PMF. 
We observe that the curves corresponding to DPA and CPA are very close to each other suggesting that DPA can achieve a high performance with lower overhead as described in Table \ref{tab:comparison}. \textcolor{black}{The main reason for this observation is that for each cell, most of the total interference is caused by a ring of neighboring cells (i.e., direct neighbors), hence, it is sufficient to only consider the direct neighbors in optimizing power coefficients for each cell as in DPA. We observed similar trends in the simulations when considering NMF or other cell radii (e.g. 500 m), but the results are omitted due to the space limit.}

\begin{figure}[h]
    \centering
        \begin{subfigure}[b]{0.49\textwidth}
            \includegraphics[width=\textwidth]{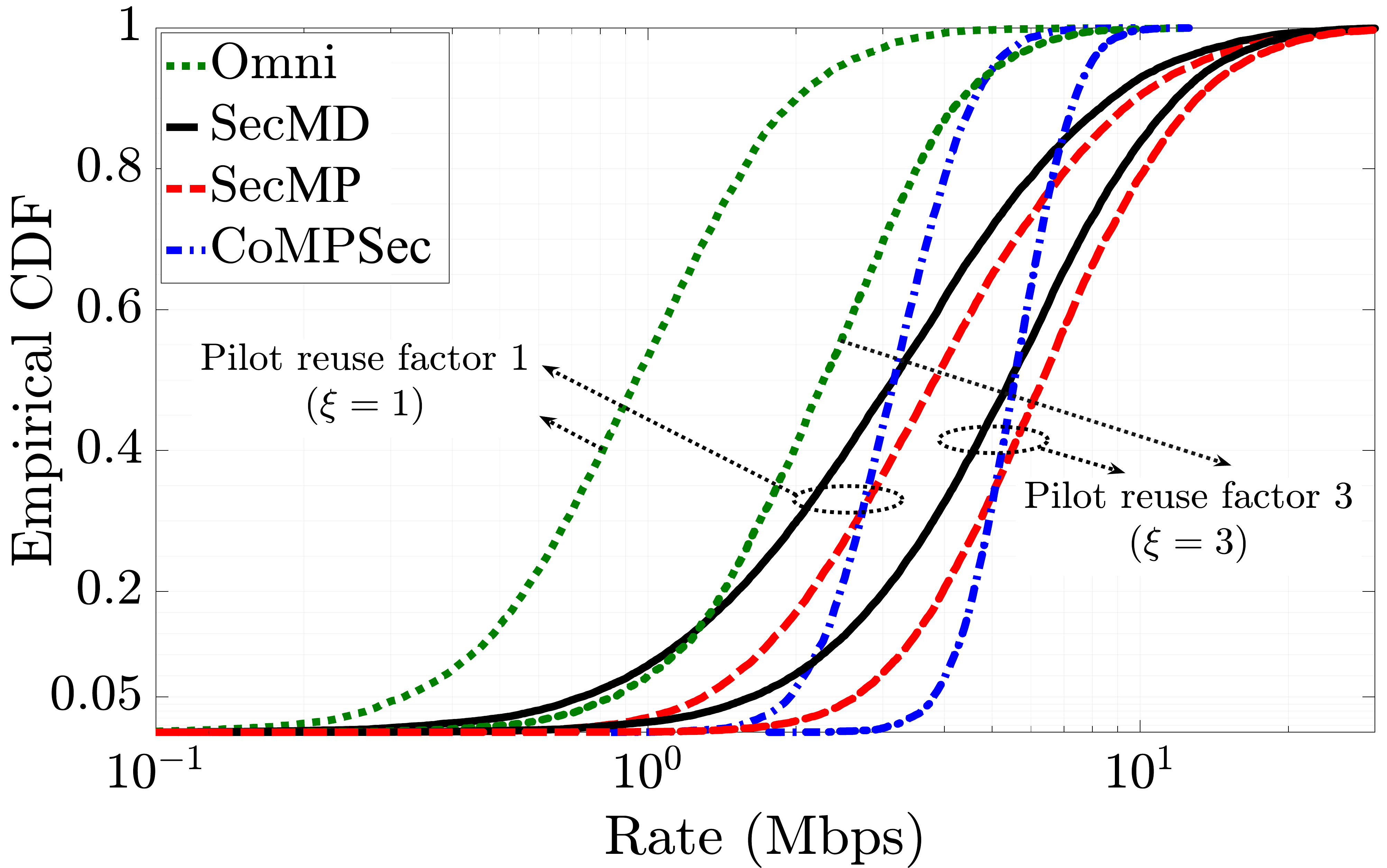}
            \caption{Maximum-ratio precoding}%
            {{\small }}    
            \label{fig:rateCDF-reuse1}
        \end{subfigure}
        \begin{subfigure}[b]{0.49\textwidth}              \includegraphics[width=\textwidth]{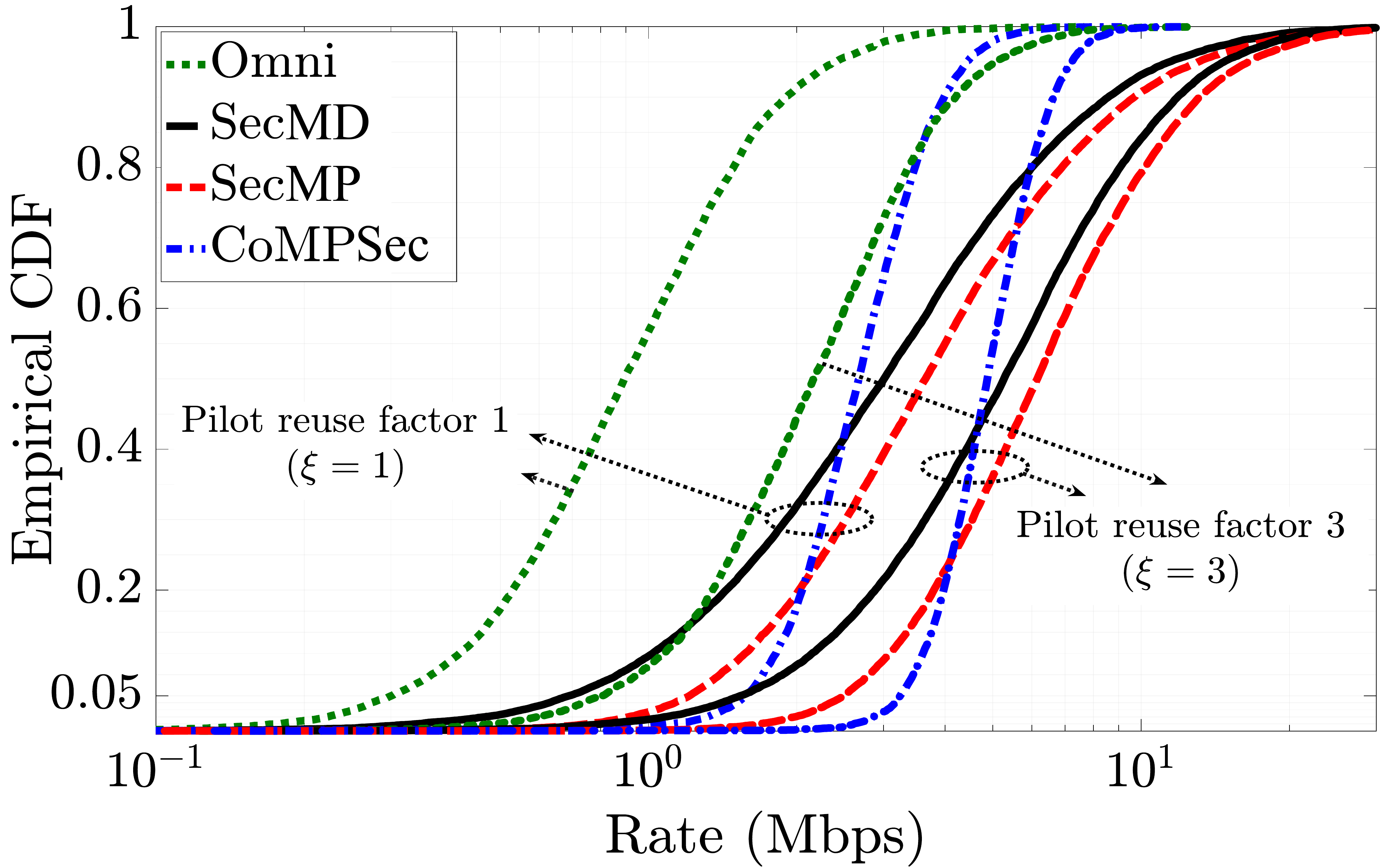}
            \caption{Zero-forcing precoding}%
            {{\small}}    
            \label{fig:rateCDF-reuse2}
        \end{subfigure}
        
        \caption{The effect of pilot reuse factor on the achievable rate when decentralized power allocation and per-cell max-min fairness are considered.}
        \label{fig:rateCDF-reuse}
         
\end{figure}

The effect of pilot reuse factor on the achievable rate is examined in Figure \ref{fig:rateCDF-reuse} where DPA is considered with PMF. We can see that a higher performance is achieved with pilot reuse factor $\xi=3$ compared to $\xi=1$. As explained in Section \ref{subsec:sec-effect}, increasing the pilot reuse factor reduces the number of users assigned the same pilot sequence and subsequently enhances the channel estimation quality. On the other hand, The cost of higher pilot reuse factor is that we need to assign larger number of resource blocks to channel estimation ($\tau_{\text{p}}=\xi K$) which remains smaller number of resource blocks for data transmission. However, we have $K=18$ users per cell and $\tau_{\text{c}}=2$ (msec) $\times~210$ (KHz) $=420$ samples. As $K<<\tau_{\text{c}}$, the benefits of increasing $\xi$ outweigh the cost in this case. \textcolor{black}{Note that increasing pilot reuse factor is not always helpful and after some point, the overhead cost outweigh the reduction of pilot contamination impact.} 

Another observation from Figure \ref{fig:rateCDF-reuse} is that sectorized settings (SecMD and SecMP) with pilot reuse factor $\xi=1$ outperform omnidirectional setting with pilot reuse factor $\xi=3$. This implies that the effect of (\textit{i}) having sectorized antennas with ideal transmission pattern is higher than (\textit{ii}) using omnidirectional setting with pilot reuse factor three instead of one. Note that in both of (\textit{i}) and (\textit{ii}) cases, the number of interfering pilots is a third compared to when we have omnidirectional setting with pilot reuse factor one. However, in case (\textit{ii}), there are fewer resource blocks for data transmission as a result of having longer pilot sequences. Furthermore, the selection diversity in SecMP setting increases the gap between (\textit{i}) and (\textit{ii}).      

\subsection{Impact of antenna pattern}
\begin{figure}[h]
    \centering
        \begin{subfigure}[b]{0.49\textwidth}
            \includegraphics[width=\textwidth]{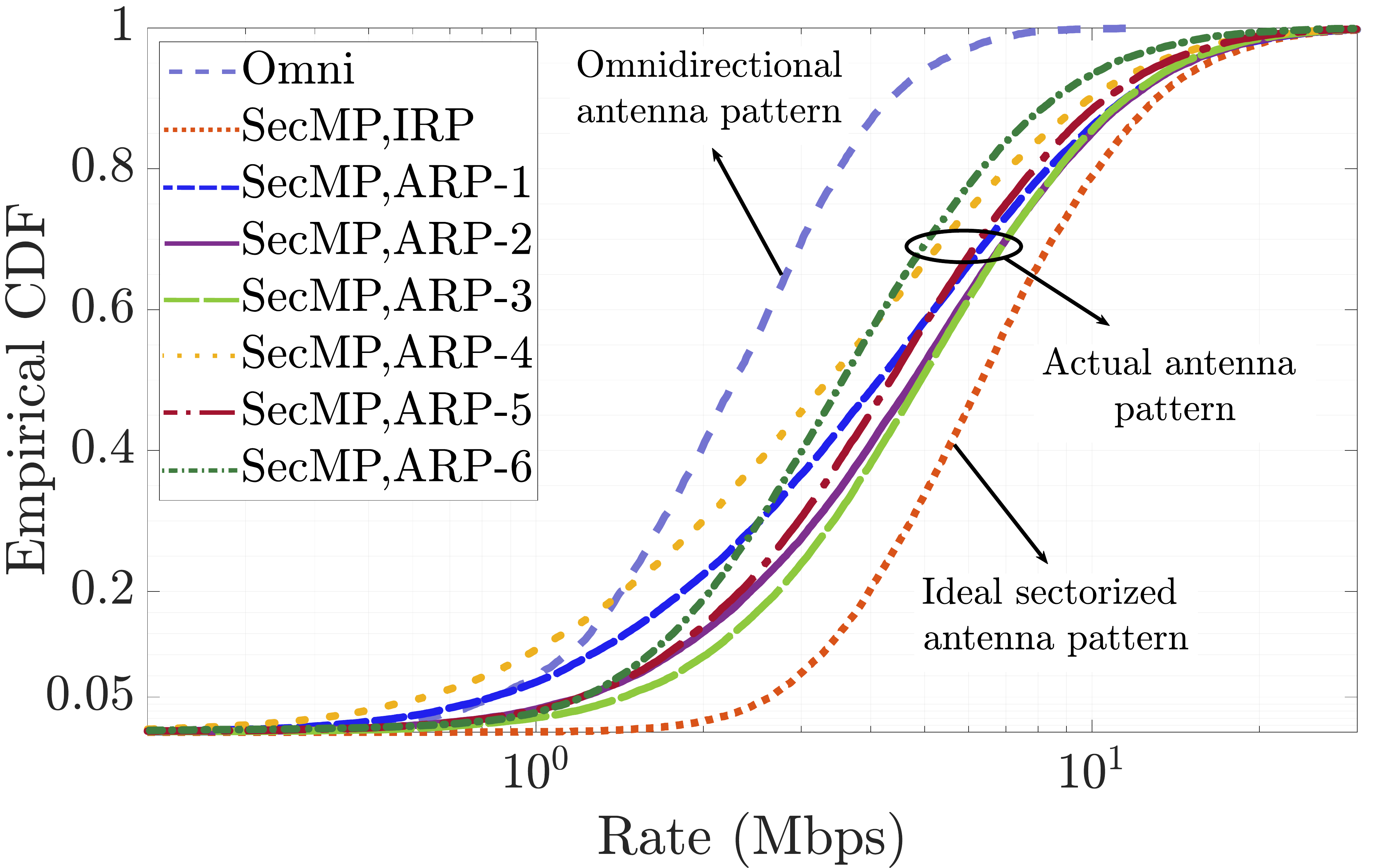}
            \caption{Maximum-ratio precoding, SecMP setting}%
            {{\small }}    
            \label{fig:rateCDF-ARP-Sec-MR}
        \end{subfigure}
        \begin{subfigure}[b]{0.49\textwidth}              \includegraphics[width=\textwidth]{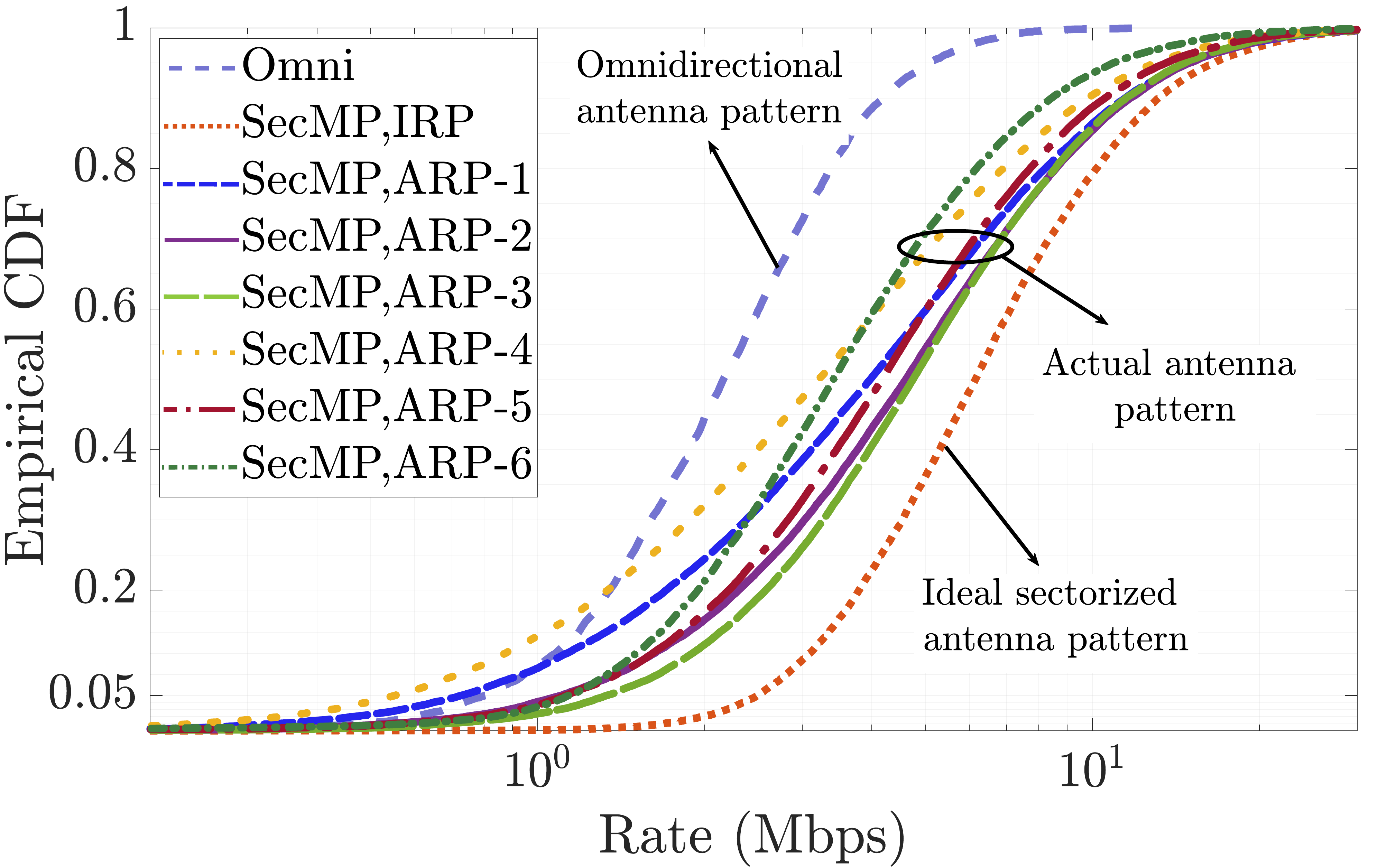}
            \caption{Zero-forcing precoding, SecMP setting}%
            {{\small}}    
            \label{fig:rateCDF-ARP-Sec-ZF}
        \end{subfigure}
        
         \begin{subfigure}[b]{0.49\textwidth}              \includegraphics[width=\textwidth]{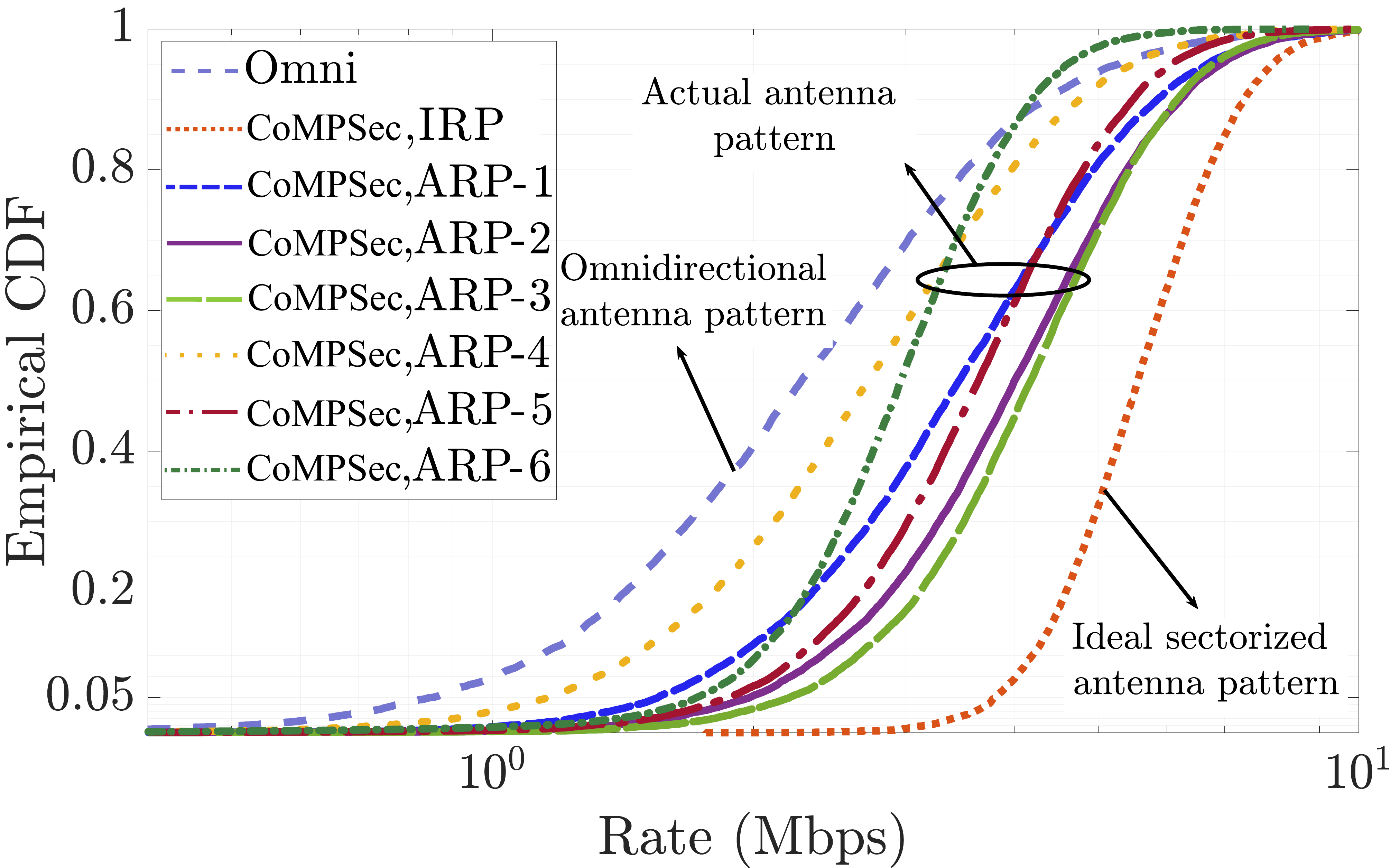}
            \caption{Maximum-ratio precoding, CoMPSec setting}%
            {{\small}}    
            \label{fig:rateCDF-ARP-Comp-MR}
        \end{subfigure}
        \begin{subfigure}[b]{0.49\textwidth}              \includegraphics[width=\textwidth]{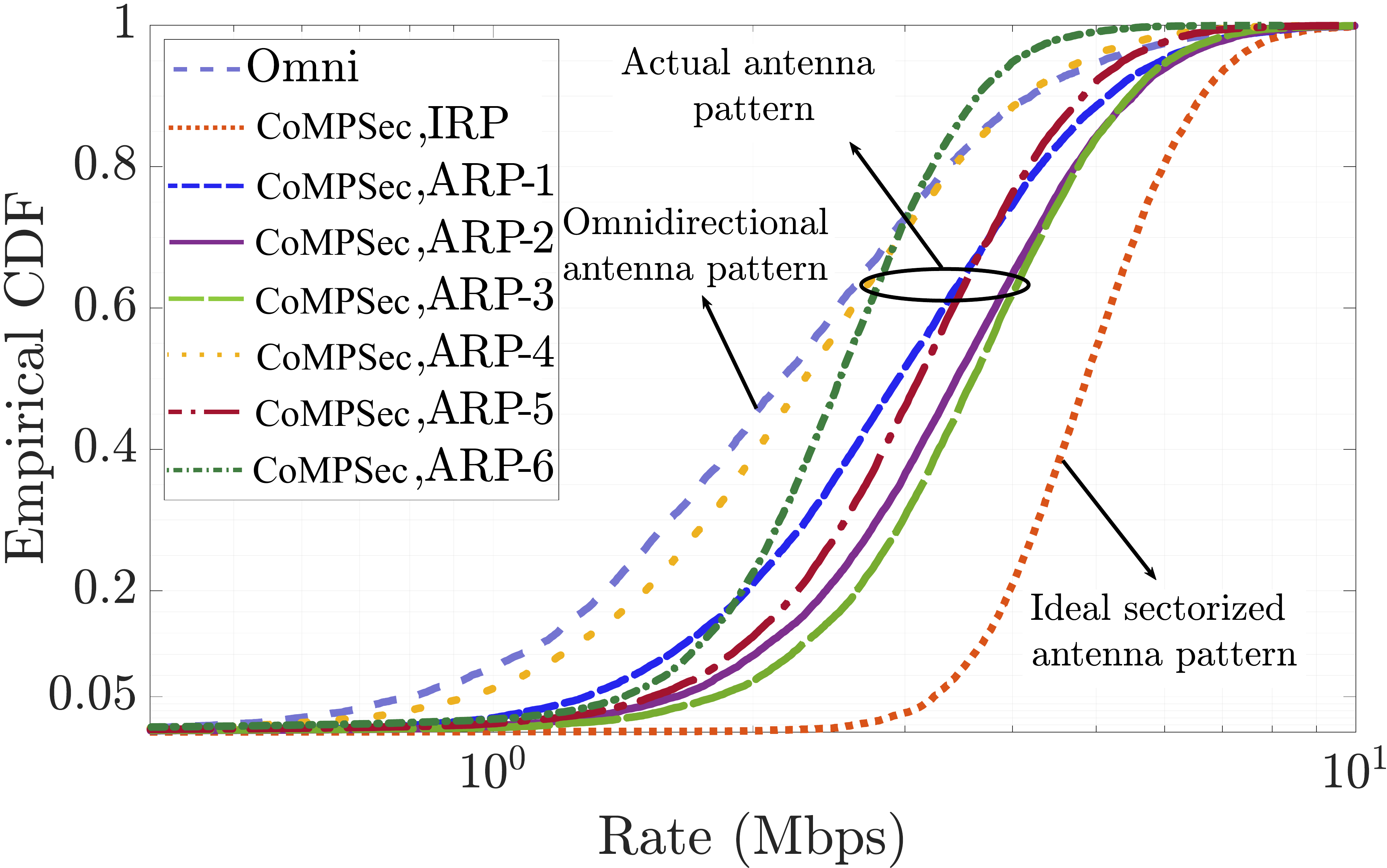}
            \caption{Zero-forcing precoding, CoMPSec setting}%
            {{\small}}    
            \label{fig:rateCDF-ARP-Comp-ZF}
        \end{subfigure}
        
        \caption{ The effect of antenna pattern on the achievable rate for different communication settings.}
        \label{fig:rateCDF-ARP}
\end{figure}

In this section, we investigate the impact of having an actual radiation pattern as opposed to an ideal pattern. To this end, we consider pilot reuse factor $\xi=3$ and a decentralized power allocation with per-cell max-min criterion. Figure \ref{fig:rateCDF-ARP} depicts the empirical CDF of the downlink achievable rate for SecMP and CoMPSec settings for the ideal radiation pattern described in Section \ref{subsec:antenna} as well as the actual radiation patterns plotted in Figure \ref{fig:patch-pattern}. As the trends are similar for both MR and ZF precoding schemes, we focus on the former. According to Table \ref{tab:arp}, the beamwidth of ARP-1 ($74^\circ$) and ARP-4 ($75^\circ$) are significantly smaller than $120^\circ$ which diminishes the performance. Furthermore, the back-lobe radiation is high in ARP-4 (see Figure \ref{fig:patch-pattern}) which worsen the performance especially for low-SINR users. The beamwidth is higher for ARP-2 ($96^\circ$) and ARP-5 ($100^\circ$) which improves the performance. The beamwidth is larger than $120^\circ$ in ARP-6 ($151^\circ$) which degrades the performance as it creates higher interference on the neighboring cells. ARP-3 leads to the best performance among the patterns as its beamwidth ($110^\circ$) is closer to $120^\circ$ and its back-lobe radiation is relatively low. 
Table \ref{tab:comparison-actual} lists the $95\%-likely$ rate as well as the median rate when considering ARP-3. By comparing the numbers in this table to the numbers in Table \ref{tab:comparison-ideal}, we can see that although an actual radiation pattern leads to less improvement than an ideal pattern, the gains are still substantial. \textcolor{black}{Additionally, we observe that the $95\%-likely$ rate improvement due to having CoMP is still significant with APR.}

\begin{table}[h]

\centering
\caption{$95\%-likely$ rate and median rate when MR precoding and DPA-PMF are applied. ARP-3 is considered. The number in parentheses is the gain compared to the benchmark setting. }
\scalebox{1}{
\begin{tabular}{ccll} \toprule
\begin{tabular}{c}\textbf{Pilot reuse}\\\textbf{factor}\end{tabular}&\begin{tabular}{c}\textbf{Communication}\\\textbf{Setting}\end{tabular} & \begin{tabular}{c}$\bf 95\%-likely$\\$\bf rate$ (Mbps)\end{tabular} & \begin{tabular}{l} $\bf Median~rate$\\~~~~(Mbps) \end{tabular} \\\toprule
$\xi=1$ & \begin{tabular}{l} \textbf{Omni}\\ \textbf{SecMD}\\\textbf{SecMP}\\\textbf{CoMP} \end{tabular} & \begin{tabular}{l} 0.31 (benchmark)\\ 0.36 ($1.16\times$)\\0.55 ($1.77\times$)\\0.85 ($2.74\times$) \end{tabular} &\begin{tabular}{l} 0.98 (benchmark)\\ 2.23 ($2.27\times$)\\2.68 ($2.73\times$)\\2.03 ($2.07\times$) \end{tabular}\\ 
\bottomrule
$\xi=3$ & \begin{tabular}{c} \textbf{Omni}\\ \textbf{SecMD}\\\textbf{SecMP}\\\textbf{CoMP} \end{tabular}		 & \begin{tabular}{l} 0.84 (benchmark)\\ 0.87 ($1.03\times$)\\1.42 ($1.69\times$)\\2.20 ($2.61\times$) \end{tabular} & \begin{tabular}{l} 2.38 (benchmark)\\ 4.17 ($1.75\times$)\\4.92 ($2.06\times$)\\4.17 ($1.75\times$) \end{tabular}   \\ 
\bottomrule
\end{tabular}
}
\label{tab:comparison-actual}
\end{table}


\section{Conclusion} \label{sec:conclusion}
In this paper, we have studied the impact of three-fold sectorization and multi-point coordination with and without sectorization on the performance of massive MIMO systems. We have shown that using sectorized antenna elements mitigates the impact of pilot contamination by attenuating the interfering pilots received from the back-lobe region of the antennas during channel estimation. We have provided downlink rate analysis and various power control strategies for both maximum-ratio and zero-forcing processing schemes. 
Simulation results have revealed that using sectorized antennas combined with multi-point coordination improves the $95\%-likely$ rate substantially. A natural extension to this work is to consider more complicated channel models to capture the spatial correlation among user channels. Another avenue for future research is to consider more complicated precoding schemes such as single-cell and multi-cell MMSE.

\IEEEpeerreviewmaketitle

\appendix

\subsection{Proof of Theorem \ref{thm:zf}}
\label{app:thm:zf}

As seen in Section \ref{subsec:zf}, when ZF precoding is applied at each antenna array the signal received by $U_{lk}$ is $y^{\text{zf}}_{lk}= T^{\text{zf}}_{11}+T^{\text{zf}}_{12}+T_2+T_3+T_4+T_5$, where $T^{\text{zf}}_{11}$ is the desired signal and other terms are interference and noise. Furthermore, it can be shown that any pair of these terms are uncorrelated. According to Theorem 1 of \cite{hassibi}, the worst case of uncorrelated additive noise is independent Gaussian
noise with the same variance. Therefore, given ZF precoding at each antenna array, $U_{lk}$'s downlink rate, $R^{\text{zf}}_{lk}$, is lower bounded by $ \frac{\tau_{\text{dl}}}{\tau_{\text{c}}} I \left(y^{\text{zf}}_{lk};q_{lk} \big| g^{\text{eff,zf}}_{lk}  \right)
 \geq \frac{\tau_{\text{dl}}}{\tau_{\text{c}}} W\log_2 \left(1+\text{SINR}^{\text{zf}}_{lk}\right)$, 

\begin{align}
\text{SINR}^{\text{zf}}_{lk}  = \Var{T^{\text{zf}}_{11}}\Big/\left(\Var{T^{\text{zf}}_{12}}+\Var{T_2}+\Var{T_3}+\Var{T_4}+\Var{T_5}\right) \label{SINR-zf}
\end{align}
The variance of the terms  $T_2$, $T_3$, $T_4$, and $T_5$ are given by \eqref{eq:T2-T5}. Furthermore we have 
\begin{align*}
    \Var{T^{\text{zf}}_{11}}=\rho_{dl}\left(\sum_{i\in [3]} e_{lik}\sqrt{\left(M-K_{li}\right) a_{lk}^{li} \gamma_{lk}^{li}\eta_{lik}}\right)^2, 
    \Var{T^{\text{zf}}_{12}}= \rho_{dl}\sum_{l'\in P_l\textbackslash\{l\}}\left(\sum_{i\in [3]} e_{l'ik}\sqrt{\left(M-K_{l'i}\right)a_{lk}^{l'i}\gamma_{lk}^{l'i}\eta_{l'ik}}\right)^2 
\end{align*}
Substituting these terms in (\ref{SINR-zf}) will conclude the proof.

\subsection{Proof of Theorem \ref{thm:mr}}
\label{app:thm:mr}
The proof is similar to the proof of Theorem \ref{thm:zf}. Applying MR precoding at each antenna array, the signal received by $U_{lk}$ is $y^{\text{mr}}_{lk}= T^{\text{mr}}_{11}+T^{\text{mr}}_{12}+T^{\text{mr}}_{13}+T^{\text{mr}}_{14}+T_2+T_3+T_4+T_5$, where $T^{\text{mr}}_{11}$ is the desired signal and other terms are interference and noise. It can be shown that any pair of these terms are uncorrelated. According to Theorem 1 of \cite{hassibi}, the worst case of uncorrelated additive noise is independent Gaussian noise with the same variance. Therefore, given MR precoding at each antenna array, $U_{lk}$'s downlink rate, $R^{\text{mr}}_{lk}$, is lower bounded by $\frac{\tau_{\text{dl}}}{\tau_{\text{c}}}I \left(y^{\text{mr}}_{lk};q_{lk} \big| g^{\text{eff,mr}}_{lk}  \right)
 \geq \frac{\tau_{\text{dl}}}{\tau_{\text{c}}} W\log_2 \left(1+\text{SINR}^{\text{mr}}_{lk}\right)$, where
\begin{align}
\text{SINR}^{\text{mr}}_{lk}  = \Var{T^{\text{mr}}_{11}}/\left(\Var{T^{\text{mr}}_{12}}+\Var{T^{\text{mr}}_{13}}+\Var{T^{\text{mr}}_{14}}+\Var{T_2}+\Var{T_3}+\Var{T_4}+\Var{T_5}\right).\notag 
\end{align}
The variance of the terms  $T_2$, $T_3$, $T_4$, and $T_5$ are given by \eqref{eq:T2-T5}. Furthermore we have
\begin{align}
    &\Var{T^{\text{mr}}_{11}}=\rho_{dl}\bigg(\sum_{i\in [3]} e_{lik}\sqrt{M a_{lk}^{li} \gamma_{lk}^{li}\eta_{lik}}\bigg)^2, 
    &&\Var{T^{\text{mr}}_{12}}= \rho_{dl}\sum_{l'\in P_l\textbackslash\{l\}}\bigg(\sum_{i\in [3]} e_{l'ik}\sqrt{M a_{lk}^{l'i}\gamma_{lk}^{l'i}\eta_{l'ik}}\bigg)^2. \label{eq:T12-mr}\\
    &\Var{T^{\text{mr}}_{13}}= \rho_{dl}\sum_{l'\in P_l}\sum_{i\in [3]} e_{l'ik}a_{lk}^{l'i}\gamma_{lk}^{l'i}\eta_{l'ik}, 
    &&\Var{T^{\text{mr}}_{14}}= \rho_{dl}\sum_{l'\in P_l}\sum_{i\in [3]} e_{l'ik}a_{lk}^{l'i}\gamma_{lk}^{l'i}\bigg(\sum_{k'\in \mathcal{A}_{l'i}}\eta_{l'ik'}\bigg). \label{eq:T14-mr}
\end{align}

\subsection{Proof of Lemma \ref{lem:quasiconcavity}}
\label{app:lem:quasiconcavity}

The constraints of problem \eqref{max-min-sinr-comp} are convex. To prove the quasi-concavity, it suffices to show that the objective function in \eqref{max-min-sinr-comp} is quasi-concave. Define $\Omega \triangleq \{\psi_{lik},X_{lk},Y_{lk}\}$ for $(l,i,k)\in [L]\times[3]\times[K]$, the set of optimization variables. In \eqref{max-min-sinr-comp}, the objective function is
$f(\Omega)=\min_{l,k} \frac{\left( \sum_{i\in[3]} J^{\text{s}}_{lik} \psi_{lik} \right)^2}{ X_{lk}^2 +  Y_{lk}^2 + Z_{lk}^2+ 1}. \notag$
%
For every $\rho \geq 0$, the upper-level set of $f(\Omega)$ is $U(f,\rho)=\left\{\Omega: f(\Omega) \geq \rho\right\}$ which is equal to
$\left\{\Omega: \frac{\left( \sum_{i\in[3]} J^{\text{s}}_{lik} \psi_{lik} \right)^2}{ X_{lk}^2 +  Y_{lk}^2 + Z_{lk}^2+ 1} \geq \rho, \forall (l,k)\right\} 
= \left\{\Omega:||V_{lk}|| \leq \frac{1}{\sqrt{\rho}} \sum_{i\in[3]} J^{\text{s}}_{lik} \psi_{lik}, \forall (l,k)\right\}, \notag$
where $V_{lk} \triangleq [X_{lk},Y_{lk},Z_{lk},1]$. The upper-level set $U(f,\rho)$ is convex since it can be represented as a second order cone. Thus, $f(\Omega)$ is quasi-concave. Note that the argument is valid for $\text{s}\in\{\text{zf},\text{mr}\}$. 

\subsection{Proof of Lemma \ref{lem:comp-pmf}}
\label{app:lem:sec-pmf}
We define $f^{\text{s}}_{lk}(\eta_{lk})\triangleq \frac{ r^{\text{s}}_{lk}\eta_{lk} }{  \frac{1}{3}\sum_{l'\in P_l} \sum_{i\in [3]} n^{l'i,\text{s}}_{lk} +  \frac{1}{3}\sum_{l'\not\in P_l} \sum_{i\in [3]} o^{l'i,\text{s}}_{lk} + 1}$. In \eqref{max-min-sinr-comp-local2}, the goal is to maximize the minimum of $f^{\text{s}}_{lk}(\eta_{lk})$ over $l$ and $k$. We note that the denominator of $f^{\text{s}}_{lk}(\eta_{lk})$ is not a function of power coefficients and the numerator is proportional to $\eta_{lk}$. We claim that $f^{\text{s}}_{lk}(\eta^{*,\text{s}}_{lk})=f^{\text{s}}_{lk'}(\eta^{*,\text{s}}_{lk'})$, $\forall k,k'$. To prove the claim, we use contradiction. Assume that the claim is incorrect. Consequently, there exists non-empty set of users $\mathcal{K}$ for which $f_{lk}(\eta_{lk})$ takes its minimum value. Thus for every $k''\in \mathcal{K}$ we have $f^{\text{s}}_{lk''}(\eta^{*,\text{s}}_{lk''})<f^{\text{s}}_{lk}(\eta^{*,\text{s}}_{lk})$, $\forall k\not\in \mathcal{K}$. Let $\eta^{\text{s}}_{lk''}=\eta^{*,\text{s}}_{lk''}+\frac{\epsilon}{|\mathcal{K}|}$, $k'' \in \mathcal{K}$ and $\eta^{\text{s}}_{lk}=\eta^{*,\text{s}}_{lk}-\frac{\epsilon}{K-|\mathcal{K}|}$, $\forall k\not\in \mathcal{K}$. It is straightforward to see that $\left\{\eta^{\text{s}}_{lk}\right\}_{k\in [K]}$ is a valid power allocation satisfying the constraints of problem \eqref{max-min-sinr-comp-local2}. As $f_{lk}(\eta_{lk})$ is proportional to $\eta_{lk}$, we can always find an $\epsilon>0$ such that the $\min_k\left(f_{lk}(\eta^{\text{s}}_{lk})\right)>\min_k\left(f_{lk}(\eta^{*,\text{s}}_{lk})\right)$ which is in contradiction with the optimality of power allocation $\left\{\eta^{*,\text{s}}_{lk}\right\}_{k\in [K]}$. Therefore, the claim is correct and we have $\overline{\text{SINR}}^{\text{s}}_l= f^{\text{s}}_{lk}(\eta^{*,\text{s}}_{lk}), \forall k$. Solving this equation for $\eta^{*,\text{s}}_{lk}$ we have $\eta^{*,\text{s}}_{lk}=\frac{D^{\text{s}}_{lk}}{r^{\text{s}}_{lk}}~\overline{\text{SINR}}^{\text{s}}_l$ where $D^{\text{s}}_{lk}$ is provided by Lemma \ref{lem:comp-pmf}. Substituting this expression into the power constraint of problem \eqref{max-min-sinr-comp-local2} (i.e. $\sum_k\eta^{*,\text{s}}_{lk}=1/3$ ), one can find the closed-form expression of $\overline{\text{SINR}}^{\text{s}}_l$ as expressed in the statement of Lemma \ref{lem:comp-pmf}.

\newcommand*{\bibfont}{\footnotesize}
\bibliographystyle{IEEEtran}
\bibliography{directional}
\end{document}